\documentclass{article}

\usepackage{arxiv}
\pdfoutput=1 

\usepackage[utf8]{inputenc} 
\usepackage[T1]{fontenc}    
\usepackage{hyperref}       
\usepackage{url}            
\usepackage{booktabs}       
\usepackage{amsfonts}       
\usepackage{nicefrac}       
\usepackage{microtype}      
\usepackage{graphicx}
\usepackage[round]{natbib}
\usepackage{doi}
\usepackage{amsmath}
\usepackage{longtable}      
\usepackage{array}          
\usepackage{dcolumn}        
\usepackage{multirow}
\usepackage{rotating}

\usepackage{caption}
\title{High Socioeconomic Status is Associated with Diverse Consumption across Brands and Price Levels}

\author{
	\hspace{1mm}\small{Yuanmo He} \\
	\small{Department of Methodology}\\
	\small{London School of Economics and Political Science}\\
	\small{Houghton Street}\\
	\small{London, WC2A 2AE, United Kingdom} \\
	\small{\texttt{\href{mailto:y.he54@lse.ac.uk}{y.he54@lse.ac.uk}}} \\
	\And
	\hspace{1mm}\small{Milena Tsvetkova} \\
	\small{Department of Methodology}\\
	\small{London School of Economics and Political Science}\\
	\small{Houghton Street}\\
	\small{London, WC2A 2AE, United Kingdom} \\
	\small{\texttt{\href{mailto:m.tsvetkova@lse.ac.uk}{m.tsvetkova@lse.ac.uk}}} \\
}

\renewcommand{\shorttitle}{High SES is Associated with Diverse Consumption across Brands and Price Levels}

\hypersetup{
pdftitle={High Socioeconomic Status is Associated with Diverse Consumption across Brands and Price Levels},
pdfsubject={Sociology, Social Psychology, Consumer Behaviour, Computers and Society},
pdfauthor={Yuanmo He, Milena Tsvetkova},
pdfkeywords={consumption behaviour, consumer brands, socioeconomic status, inequality, human mobility data},
}

\begin{document}
\maketitle

\begin{abstract}
\frenchspacing Consumption practices are determined by a combination of economic, social, and cultural forces. We posit that lower economic constraints leave more room to diversify consumption along cultural and social aspects in the form of omnivorous or lifestyle-based niche consumption. We provide empirical evidence for this diversity hypothesis by analysing millions of mobile-tracked visits from thousands of Census Block Groups to thousands of stores in New York State. The results show that high income is significantly associated with diverse consumption across brands and price levels. The associations between diversity and income persist but are less prominent for necessity-based consumption and for the densely populated and demographically diverse New York City. The associations replicate for education as an alternative measure of socioeconomic status and for the state of Texas. We further illustrate that the associations cannot be explained by simple geographic constraints, including the neighbourhoods' demographic diversity, the residents' geographic mobility and the stores' local availability, so deeper social and cultural factors must be at play.
\end{abstract}


\section{Introduction}

An important aspect of socioeconomic inequality is the difference in daily consumption practices by socioeconomic status (SES). This difference is not only a manifestation of inequality, but also a trigger for further inequality in other life outcomes. For example, constrained by availability and price, people in low SES tend to go to low-price supermarkets and consume unhealthy food and beverages, which could contribute to later health problems \citep{pechey_supermarket_2015, zagorsky_association_2017, baumann_food_2019}. Also, in more unequal societies, people tend to pay more attention to and spend more money on status goods \citep{heffetz_test_2011, walasek_income_2015, walasek_positional_2018}, which may make them work more \citep{bowles_emulation_2005}, save less \citep{wisman_household_2009}, accrue more debt \citep{christen_keeping_2005}, or even declare bankruptcy \citep{perugini_inequality_2016}.

Consumption is an economic, social, and cultural phenomenon that attracts research from multiple disciplines. From an economic perspective, consumption inequality is an important complement to income inequality, as the basic utility function of individuals typically includes consumption and leisure, and not necessarily income \citep{attanasio_consumption_2016}. From the angle of social psychology, consumption practices express personal and social identities \citep{belk_possessions_1988, woodward_divergent_2003, reimer_identity_2004} and social comparison and status seeking drive people to engage in consumption practices with the sole purpose to demonstrate how successful and rich they are \citep{sahin_socioeconomic_2022, walasek_positional_2018, walasek_income_2015}. From the view of cultural sociology, people's consumption of music, theatre, and other cultural events encode their socioeconomic background and serve to distinguish their social position; thus, cultural consumption both reflects and reproduces socioeconomic inequality \citep{bourdieu_distinction_1984}.

In this paper, we integrate insights from sociology, social psychology, and consumer research and present new large-scale empirical evidence for socioeconomic inequality in daily consumption. We argue that lower economic constraints make daily consumption practices more focused on social and cultural distinction and thus, higher SES individuals engage in more omnivorous \citep{peterson_understanding_1992, peterson_changing_1996} as well as more niche consumption \citep{berger_subtle_2010, eckhardt_rise_2015}. As a result, higher SES is associated with more diverse consumption practices. We analyse mobile tracking data of New York State residents' visits to the stores of various brands to present evidence for this hypothesis. Our findings illustrate and quantify socioeconomic divisions in daily consumption practices, bearing further evidence for the pervasiveness and inevitability of inequality in daily life.

\section{Background}

Although it is widely accepted that consumption behaviour is associated with SES, both the nature of the relationship and our understanding of it continue to evolve. Classical sociological theories emphasize the role of consumption for socioeconomic affirmation and differentiation. Veblen's ([1899] \citeyear{veblen_theory_2017}) notion of \emph{conspicuous consumption} captures the idea that some consumption practices are not economically practical but strategically employed to impress or show off to others. Bourdieu's \citeyearpar{bourdieu_distinction_1984} theory of cultural capital and taste indicates that people engage in highbrow cultural consumption practices to distinguish higher social status. However, more recent research has challenged the idea that elites rely on conspicuous and highbrow consumption alone. Contemporary high-SES individuals tend to be omnivorous in their cultural consumption, engaging with various cultural genres instead of focusing solely on highbrow practices \citep{peterson_understanding_1992, chan_understanding_2019, de_vries_what_2021}. Some high-SES people, especially those with higher cultural capital, also tend to enjoy inconspicuous and niche consumption practices \citep{berger_subtle_2010, eckhardt_rise_2015, eckhardt_new_2020}.

While these theories and findings may appear contradictory, we argue that they point to a shared underlying pattern: social and cultural factors dominate consumption mainly when economic constraints are weak. For high-SES individuals, consumption practices are less driven by product prices and more influenced by social and cultural processes of distinction. The omnivore and niche preferences are simply two different forms of status distinction for the well-off.

On the one hand, distinction could occur in reference to low SES. For people in high SES, high economic capital removes constraints of resources, high cultural capital provides the ability to be open-minded and appreciate diversity, and high social capital offers wider exposure to various consumption practices \citep{bourdieu_distinction_1984, peterson_understanding_1992, peterson_changing_1996, chan_understanding_2019}. These conditions have been used to explain cultural omnivorousness, but they still hold true for daily goods and services. Hence, to display broad-mindedness, progressiveness, and non-materialism, some high-SES individuals may engage in omnivorous consumption practices, consuming across the spectrum of brands and prices.

On the other hand, distinction may be confined within the high-SES stratum. People of high SES have the capacity to use consumption to demonstrate dedication to a lifestyle or ideology such as luxury and exclusivity, but also environmental sustainability, healthy and natural living, New Age beliefs, anti-globalization, even anti-consumerism in the form of inconspicuous, experiential, or authenticity consumption \citep{eckhardt_new_2020}. In contrast, low-SES individuals who pinch pennies to provide bare essentials such as food, housing, and fuel do not have this privilege. Thus, to affirm their identity and lifestyle, some high-SES individuals may consume within market niches, bundling brands and services with intention and purpose.

\begin{figure}[ht]
\centering
\includegraphics[width=0.9\textwidth]{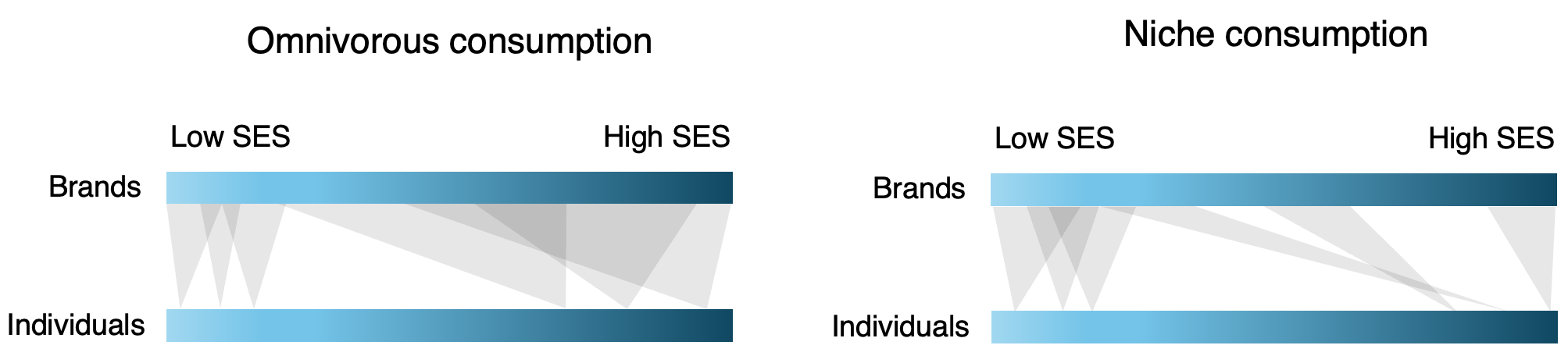}
\caption{Both omnivorous and niche consumption hypotheses imply high SES associates with more diverse consumption.}
\label{fig:fig1}
\end{figure}

While omnivorousness implies that high-SES individuals engage in diverse consumption practices, niche consumption implies that high-SES individuals consume within narrow niches but that there is a diversity of niches (Figure \ref{fig:fig1}). Whether taken individually or together, both processes imply that, in the aggregate, high SES is associated with more diverse consumption practices. We call this the diversity hypothesis: higher SES is associated with more diverse consumption patterns in terms of brands and price levels.

Prior research already provides partial evidence in support of omnivorous and niche consumption among high-SES individuals. The qualitative research on inconspicuous consumption among elites we already referred to corroborates the niche consumption argument \citep{berger_subtle_2010, eckhardt_rise_2015, eckhardt_new_2020}. Meanwhile, computer scientists are starting to discover evidence of the associations between SES and consumption diversity in large-scale digital trace data. Linking mobile phone with banking transaction data from Mexico and Turkey, researchers found that high SES is correlated with more diverse purchases across product and service categories \citep{leo_correlations_2018} and merchants \citep{dong_purchase_2020}, offering evidence for omnivorousness in material consumption. Yet, few studies have connected the theories and large-scale empirical evidence together to provide a coherent explanation of the phenomenon.

This paper aims to bridge the theory-driven and data-driven approaches. We extend existing work in three ways. First, we generalize the association between SES and diversity in consumption to encompass omnivorousness, niche, and inconspicuous consumption. On the one hand, we extend the omnivorousness argument from the consumption of cultural products to daily goods and services and on the other, we bring attention to the problem of inequality to studies of lifestyle, niche, and inconspicuous consumption. Second, we replicate the previous empirical findings on product categories from Mexico and Turkey for brands in the US, confirming the universal nature and broad reach of the problem we study. Third, we provide evidence with data that are easily available for researchers at a reasonable fee, in contrast to the restricted, sensitive, and private individual banking and communication data. Our data are at the aggregate and not the individual level, but they are easily accessible, allowing the wider scientific community to replicate and extend the findings presented here.

\section{Methods}

\subsection{Data}

The data for this paper come from three sources: SafeGraph, the US Census Bureau, and Yelp. SafeGraph is a company that curates geospatial data linked with mobile tracking data for a large panel of US smartphones \citep{safegraph_welcome_2022}. In the main datasets from SafeGraph, each observation is a place that is a point of interest (POI), namely, a specific physical location that people find interesting (restaurants, retail stores, grocery stores, etc.). For each POI, SafeGraph provides a range of information including the store's name, street address, counts of visits, the home census block groups (CBGs) of the visitors, the store's North American Industry Classification System (NAICS) category, the brand of the store, and other brands that the visitors visited on the same day/week/month, etc. SafeGraph provides data on visits to the stores of more than 7,000 distinct brands, covering all the major brands in the US. Around 80\% of POIs do not have an associated brand, as they are unique commercial locations (local restaurants, museums, etc.). In this paper, we only work with POIs that are associated with brands and aggregate data of POIs to the brand level.

The NAICS codes are hierarchical six-digit codes used to represent the industries, where the first two digits represent the most general categories, and the full six digits represent the most specific categories. We select 28 four-digit categories that are related to daily consumption and these contain 78 six-digit categories. We also filter by location and time. We present results based on data from New York State in October 2019. We chose October 2019 because it is the most recent month before the COVID-19 pandemic that is not affected by holiday shopping seasons. We select to focus on one state rather than the entire country since many brands are regional. New York State is a reasonable choice because it is populous, ranking fourth among US states with a population of nearly 20 million, and the most socioeconomically unequal, with income Gini coefficient of 0.51 (US Census Bureau, \citeyear{us_census_bureau_american_2022}). This guarantees enough observations and variability in the data. We have no reason to expect that the phenomenon and mechanisms we investigate here differ qualitatively for other states but, for robustness, we replicate the findings with data from another populous state, Texas (see \ref{app:appendix_f}).

After the filtering, we have 264,826 POIs in total.\ Table \ref{tab:tab_a1} in \ref{app:appendix_a} shows the four-digit category names, the NAICS codes, and the number of POIs for each category. Among these observations, 35,588 (about 13 percent) have brands associated with them. After dropping observations with limited information, we have 24,188 POIs that belong to 1,175 brands. Among the 28 categories, four of them (``Gambling Industries'', ``Museums, Historical Sites, and Similar Institutions'', ``Performing Arts Companies'', ``Special Food Services'') do not have any brand, so we end up with 24 categories.

For 23,536 of the 24,188 POIs, each observation has a table of the visitors' home CBGs and the number of visitors from each CBG. The CBG is the smallest geographic unit for which the US Census Bureau publishes sample data; it normally has population of 600 to 3000 people. The visitors to the 23,536 POIs are from 54,307 CBGs. SafeGraph provides census data from the US Census Bureau's American Community Survey 5-year Estimates \citep{safegraph_open_2022, us_census_bureau_american_2022}. We link the CBGs with its median household income from the 2019 census data, if available. Aggregating the data by brands, we get a bipartite network of 52,509 CBGs and 1,150 brands, where the weights are the number of visits from each CBG to each brand. We drop the brands with fewer than 100 incoming visitors and the CBGs with fewer than 100 outgoing visitors, resulting in 924 brands and 13,653 CBGs. Compared with all 52,509 CBGs, the selected 13,653 CBGs slightly oversample lower income CBGs, but this factor is not anticipated to exert any discernible influence on the results of our study (Figure \ref{fig:fig_a1}).

We used Yelp to get an indicator for the price level of the brands. Yelp is a social media platform that publishes crowd-sourced reviews of businesses. It is primarily used for restaurant reviews, but also includes reviews of other businesses such as clothing stores, department stores, and grocery stores. For many stores, Yelp provides a reference price level in the form of one to four dollar signs (\$) that indicates the average cost per person: under \$10, \$11-\$30, \$30-\$60, and above \$60. Yelp officially provides an open dataset that covers businesses in 11 metropolitan areas (Montreal, Calgary, Toronto, Pittsburgh, Charlotte, Urbana-Champaign, Phoenix, Las Vegas, Madison, and Cleveland). With text matching, we are able to match 371 brands in our sample with the Yelp Open Dataset. For the few brands that have stores with different price levels, we use the mode price level. For the rest of brands, we manually search the brands on Yelp, setting the location in New York City and other cities in the New York State. Combining the data from the Yelp Open Dataset and manual searching, we are able to find the Yelp price levels for 783 brands in our sample.

\subsection{Statistical Analyses}

We analyse the data from two angles: brands and CBGs. For each of the 924 brands, we use the median household income of the visitors' home CBG to obtain a distribution of the median household income of brand visitors. We then use the median value from the distribution to represent the typical household income of the brands' visitors, which we interpret as an indicator of the brands' SES. We compare the typical income of brands' visitors and the price level of the brands to examine the extent to which brand visits are determined by their price.

To test our main hypothesis, we estimate the correlations between the median household income of the CBGs and three measures of diversity of consumption. The first measure uses the standard deviation of the typical income of the brands' visitor to test whether people living in CBGs with higher median household income tend to visit more diverse brands in terms of SES. This measure is somewhat tautological as the brand's SES is constructed by the visitors' SES, so the interpretation should be cautious. The second and third measures use the normalized Shannon entropy for the brands and the brands' price levels, respectively. For a CBG $i$, the normalized Shannon entropy is:

\begin{equation*}
D(i) = \frac{-\sum_{j=1}^{k} p_{ij} \log(p_{ij})}{\log(k)}
\end{equation*}

where $k$ is the number of brands or price levels in our sample and $p_{ij}$ is the proportion of visits from CBG $i$ to brand/price level $j$ out of the total number of visits of $i$. For the entropy by brands, $k = 924$ and for the entropy by price level, $k = 4$. For robustness, we replicate the analyses with three additional measures of diversity (Table \ref{tab:tab_a2}). To provide indirect evidence for the supposed mechanisms, we also disaggregate the analyses by industry and for New York City versus the rest.

To reject alternative explanations for the observed effects, we use regression analyses to predict CBG diversity in consumption with median household income, while controlling for CBG residents' income variability and mobility and local brand availability. We measure income variability with the margin of error for the CBG from the census data. We measure mobility with the median distance travelled from the centre of the CBG to the stores. We measure local availability with the same measures we used for consumption diversity, but with the brands available within the CBG instead of the brands visited. We run the regression model for all data and for each industry separately, including fixed effects for CBGs located within New York City versus outside. To address the potential ecological fallacy due to the aggregate analysis, we also control for available demographic variables for the CBGs, including median age, proportion of male residents, proportion of white residents, and proportion of residents whose highest degree earned is a bachelor's degree or higher.

\subsection{Replicability}

We replicate the analyses with education as an alternative proxy for SES (\ref{app:appendix_e}). We measure education as the proportion of CBG residents who are 25 years and over whose highest educational attainment is a bachelor's degree or higher. This measure is highly correlated with an alternative measure, the average years of education, calculated as the weighted sum of the proportion of residents with different highest educational attainment multiplied by the years of education required (0.888, $p < 0.001$). Since the findings are the same, we do not report them here.

Finally, to test the external validity of our findings, we replicate the analyses with data from another populous state, Texas, taken over the same time period (reported in \ref{app:appendix_f}).

\section{Results}

\subsection{Descriptive results}

\begin{figure}[ht]
\centering
\includegraphics[width=\textwidth]{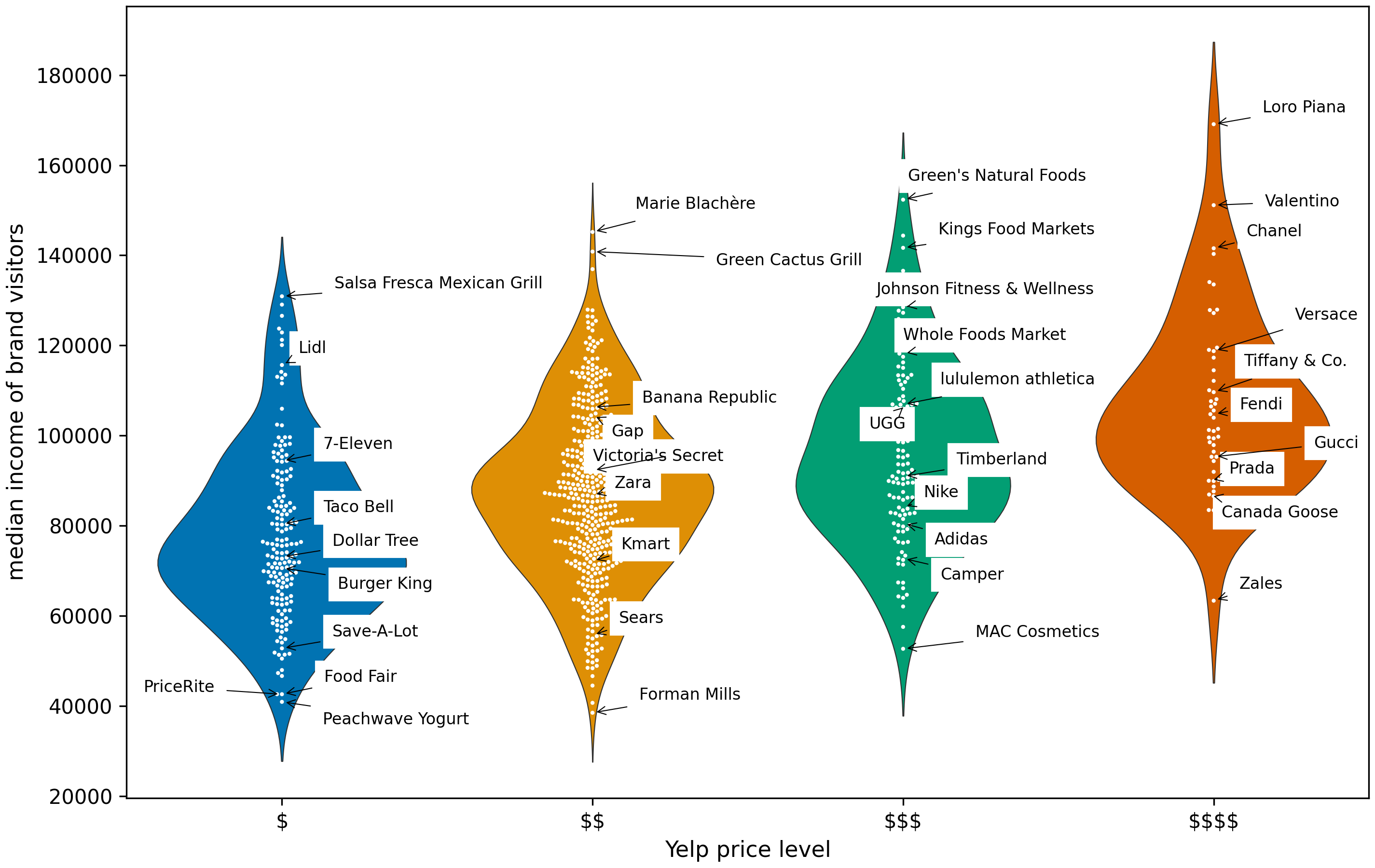}
\caption{Relation between brands' Yelp price level and median income of brand visitors' CBG. Two extreme outliers, \emph{Learning Express Toys} (median income \$203,438; Yelp price level \$\$) and \emph{Balduccis} (median income \$250,001; Yelp price level \$\$\$), are excluded for better visualisation.}
\label{fig:fig2}
\end{figure}

As expected, SES is associated with different consumption preferences for consumer brands. For instance, using a simple LASSO regression model (reported in \ref{app:appendix_b}), we can predict the median household income of the CBGs with the proportion of outgoing visitors to the 924 brands, obtaining an out-of-sample correlation between predicted and actual median CBG income of 0.748 ($p < 0.001$). This association does not necessarily correspond to economic constraints driven by the product prices. Figure \ref{fig:fig2} shows the distribution of brands' SES (measured by the median income of brand visitors) for different Yelp price levels, with some typical brands labelled. In general, pricier brands have visitors from CBGs with higher median income, but there are large variations between brands in the same price level. Some brands perform as expected. For example, discount stores such as Save-A-Lot and Price Rite have both low price levels and visitors from median income CBGs, while expensive supermarkets such as Whole Food and luxury fashion brands such as Valentino have high price levels and visitors from CBGs with high median income. Some brands show surprising patterns. For example, cheap supermarket Lidl has visitors from CBGs with relatively high median income whereas luxury fashion brand Gucci has visitors from CBGs with relatively low median income.

\begin{figure}[ht]
\centering
\includegraphics[width=\textwidth]{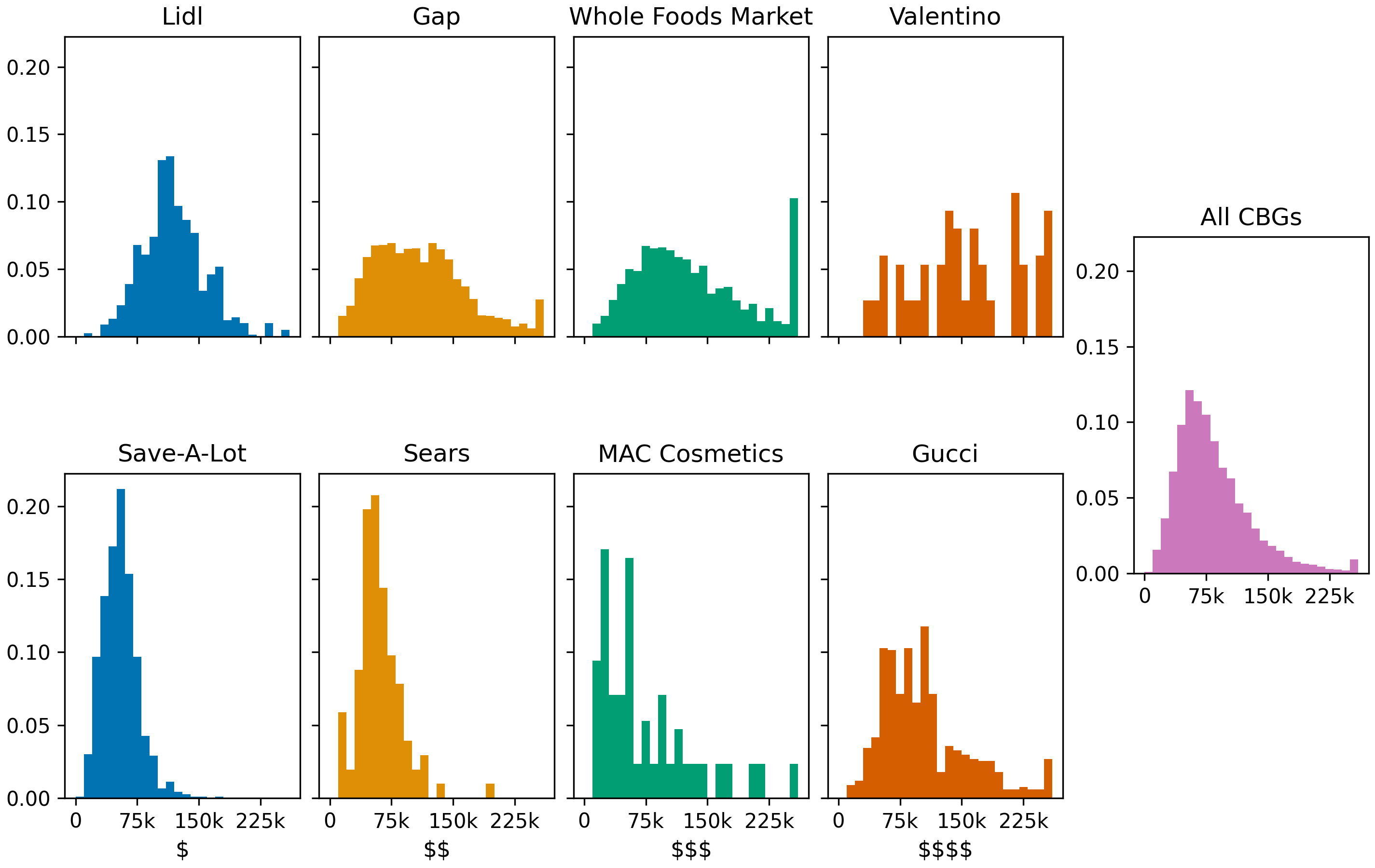}
\caption{Distribution of median household income of brand visitors' CBGs for representative brands. The distribution of median household income for all CBGs in the sample is shown for comparison.}
\label{fig:fig3}

\end{figure}

Figure \ref{fig:fig3} shows the distribution of brand visitors' CBG income for some typical brands identified from Figure \ref{fig:fig2} and, for reference, the distribution of median household income for the CBGs in our sample. As cheap grocery and department stores, Save-A-Lot and Sears have the expected distributions. But the cheap grocery and clothing stores Lidl and Gap attract large proportion of visitors from middle to high income CBGs, showing possible patterns of inconspicuous or omnivorous consumption. Whole Foods and Valentino have large proportion of visitors from middle to high income CBGs, which is expected, but they still attract low-income visitors, which may indicate conspicuous consumption. Gucci and the expensive cosmetics brand MAC Cosmetics have surprisingly large proportion of visitors from low to middle income CBGs, showing even stronger possibility of conspicuous consumption. Both luxury fashion brands, Valentino and Gucci exhibit a dramatic difference in visitors' income distribution.

\subsection{Diversity in consumption}

\begin{figure}[ht]
\centering
\includegraphics[width=\textwidth]{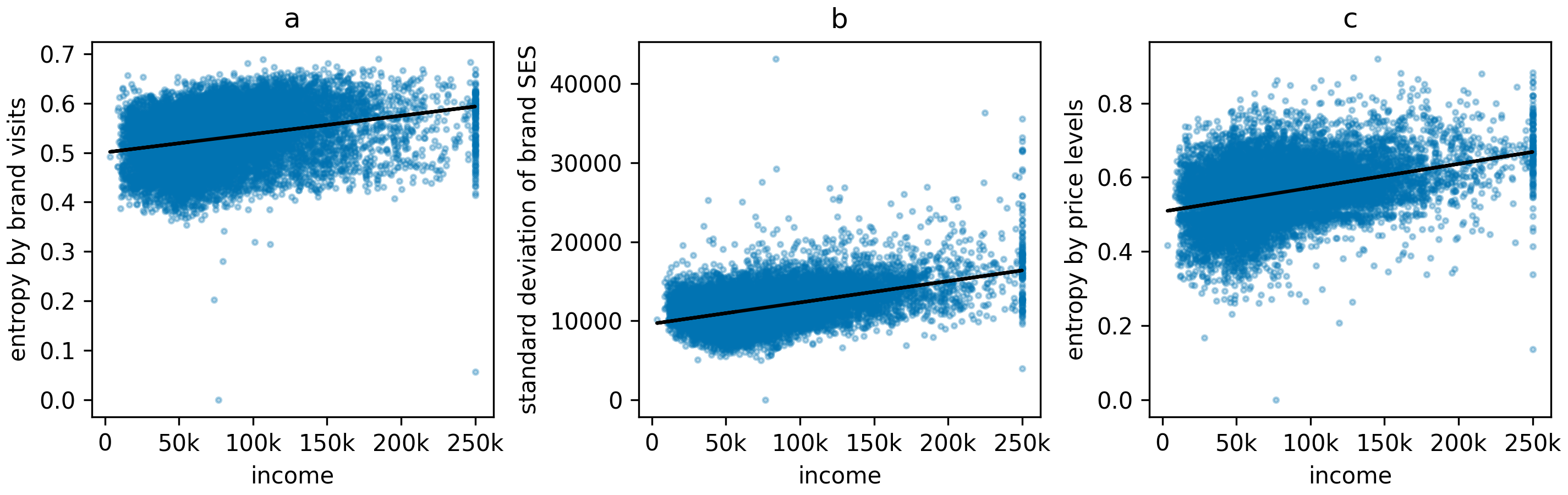}
\caption{Correlation between CBGs' median household income and consumption diversity. Results are shown for a) Shannon entropy of brand visits; b) standard deviation of brand SES; c) Shannon entropy of brand price levels.}
\label{fig:fig4}
\end{figure}

We find consistent support for the consumption diversity hypothesis. All measures of diversity have significant and positive correlation with the median household income of the visitors' CBGs: a) the Shannon entropy by brand, b) the standard deviation of visited brands' SES, and c) the Shannon entropy by brands' price level (Figure \ref{fig:fig4}). The correlation coefficients are 0.292, 0.471, and 0.358, respectively, all statistically significant at the 0.001 level. In brief, these results indicate that people residing in CBGs with higher median household income tend to visit more diverse brands, brands of more diverse SES, and brands of more diverse price levels.

To investigate how the associations between SES and diversity vary for different industries, we group similar industries together by their three-digit NAICS codes (Table \ref{tab:tab1}). There are significant associations between the median income of the brand visitors' CBGs and the diversity measures in all the industries, confirming the robustness of the diversity phenomenon. It also appears that the association is stronger in industries that involve cultural and lifestyle aspects (e.g., \emph{Food Services and Drinking Places, Clothing and Clothing Accessories Stores, Amusement, Gambling, and Recreation Industries Sporting, Goods, Hobby, Musical Instrument, and Book Stores, Miscellaneous Store Retailers}) than those that concern necessity goods (e.g., \emph{Food and Beverage Stores, Health and Personal Care Stores, General Merchandise Stores, Gasoline Stations}). The category \emph{Motion Picture and Video Industries} is an exception as we only have nine cinema chain brands here; we expect that, in reality, the diversity comes more from independent cinemas, which we do not analyse. The differences do not appear to be driven by differences in brand and price variability between industries (last three columns in Table \ref{tab:tab1}). Overall, the pattern corresponds well with the rationale of the diversity hypothesis: higher prominence of social and cultural factors should make omnivorousness and/or niche consumption more salient.

\begin{table}[ht]
\centering
\caption{The associations between CBGs' median household income and diversity in consumption by industry.}
\label{tab:tab1}
\footnotesize
\begin{tabular}{lrrrrrr}
\toprule
\textbf{Industry} & \multicolumn{3}{c}{\textbf{Associations with income}} & \multicolumn{3}{c}{\textbf{Industry characteristics}} \\
\cmidrule(lr){2-4} \cmidrule(lr){5-7}
& \textbf{Entropy by} & \textbf{Std of} & \textbf{Entropy by} & \textbf{Number of} & \textbf{Std of} & \textbf{Std of} \\
& \textbf{brand} & \textbf{brands' SES} & \textbf{brands' price} & \textbf{brands} & \textbf{brands'} & \textbf{brands'} \\
& & & \textbf{level} & & \textbf{SES} & \textbf{price level} \\
\midrule
Amusement, Gambling, and & 0.329*** & 0.246*** & 0.153*** & 89 & 19314 & 1.035 \\
Recreation Industries & & & & & & \\
Miscellaneous Store & 0.271*** & 0.216*** & 0.083*** & 49 & 23878 & 0.598 \\
Retailers & & & & & & \\
Clothing and Clothing & 0.239*** & 0.098*** & 0.232*** & 203 & 19654 & 0.768 \\
Accessories Stores & & & & & & \\
Food Services and Drinking & 0.238*** & 0.444*** & 0.404*** & 297 & 18584 & 0.663 \\
Places & & & & & & \\
Sporting Goods, Hobby, & 0.222*** & 0.254*** & 0.042*** & 53 & 25145 & 0.592 \\
Musical Instrument, and & & & & & & \\
Book Stores & & & & & & \\
Motion Picture and Video & 0.111*** & 0.074*** & -- & 9 & 21571 & -- \\
Industries & & & & & & \\
Health and Personal Care & 0.108*** & 0.022* & 0.067*** & 35 & 20154 & 0.512 \\
Stores & & & & & & \\
Gasoline Stations & 0.103*** & 0.167*** & 0.036*** & 47 & 14221 & 0.494 \\
Food and Beverage Stores & 0.070*** & 0.064*** & 0.100*** & 95 & 30776 & 0.640 \\
General Merchandise Stores & 0.050*** & 0.130*** & 0.210*** & 42 & 21078 & 0.935 \\
\bottomrule
\end{tabular}
\begin{flushleft}
\footnotesize
Note: * $p < 0.05$, ** $p < 0.01$, *** $p < 0.001$ (two-tailed tests). Industries are ordered in descending order by entropy by brand. Yelp price levels are not available for brands in \emph{Motion Picture and Video Industries}. \emph{Personal and Laundry Services} and \emph{Rental and Leasing Services} are excluded due to limited number of brands.
\end{flushleft}
\end{table}

We explore how urban-rural lifestyle differences influence consumption diversity by disaggregating the analysis for New York City (NYC) and the rest of the state (Table \ref{tab:tab2}). We study the following cases: visits from CBGs in NYC to stores in New York state, visits from CBGs outside of NYC to stores in New York state, visits from CBGs in NYC only to stores in NYC, and visits from CBGs outside of NYC to stores outside of NYC. The associations between CBG's income and the diversity measures are much stronger for people who live outside NYC than people who live in the city. This suggests that NYC dwellers engage in equally diverse consumption regardless of their income. Some of this might be explained with the high density and diversity of consumption options in the city, the weaker constraints imposed by mobility, and possibly, the stronger social comparison and influence imposed by the denser population.

\begin{table}[ht]
\centering
\caption{The associations between income and consumption diversity by CBGs in and outside New York City.}
\label{tab:tab2}
\footnotesize
\begin{tabular}{llrrr}
\toprule
\multicolumn{2}{l}{\textbf{Region}} & \textbf{Entropy by} & \textbf{Standard} & \textbf{Entropy by} \\
& & \textbf{brand} & \textbf{deviation of} & \textbf{brands' price} \\
& & & \textbf{brands' SES} & \textbf{level} \\
\midrule
\multirow{2}{*}{NYC CBGs} & visiting all stores & -0.038** & 0.134*** & 0.202*** \\
& visiting only NYC stores & 0.122*** & 0.023 & 0.041** \\
\multirow{2}{*}{Non-NYC CBGs} & visiting all stores & 0.468*** & 0.602*** & 0.530*** \\
& visiting only non-NYC stores & 0.370*** & 0.599*** & 0.506*** \\
\bottomrule
\end{tabular}
\begin{flushleft}
\footnotesize
Note: * $p < 0.05$, ** $p < 0.01$, *** $p < 0.001$ (two-tailed tests). NYC: New York City.
\end{flushleft}
\end{table}

\subsection{Robustness to alternative explanations}

The main finding that high SES is associated with diverse consumption practices could be potentially explained with geographic constraints. One such constraint is geographic mobility -- high SES individuals may have higher geographic mobility and thus access to more consumer choices. Another constraint is local availability -- high-SES CBGs may have more diverse consumer options due to urban planning or companies' strategic marketing. A third alternative explanation is that CBGs with high median income have higher variance of incomes and hence, our aggregate analyses simply capture the trivial fact that more diverse individuals consume more diversely. We find that median income is not significantly correlated with mobility and only weakly associated with local availability (0.075 for entropy by brand, 0.201 for brands' SES, 0.082 for entropy by price, $p < 0.001$ for all). Nevertheless, income is strongly correlated with income variability (0.592, $p < 0.001$), as measured with the margin of error from the census data.

We include income variability, mean distance travelled, and local availability in a regression model to predict consumption diversity by median income, including fixed effects for CBGs in NYC. To mitigate the possibility of ecological fallacy, we also control for demographic characteristics of the CBGs including median age, proportion of male residents, proportion of white residents, and proportion of residents whose highest degree earned is a bachelor's degree or higher. The variables are standardised to compare the relative importance of the predictors. Table \ref{tab:tab3} shows the results from the regression models using different diversity measures. We find that median income is significantly associated with consumption diversity even after controlling for mobility, local availability, income variability, and the demographic variables, and it has the strongest effect among the predictors.

\begin{table}[ht]
\centering
\caption{Regression results that predict CBGs' consumption diversity with median household income, controlling for confounding variables.}
\label{tab:tab3}
\footnotesize
\setlength{\tabcolsep}{6pt}
\begin{tabular}{lrrr}
\toprule
& \textbf{Entropy by} & \textbf{Standard} & \textbf{Entropy by} \\
& \textbf{brand} & \textbf{deviation of} & \textbf{brands' price} \\
& & \textbf{brands' SES} & \textbf{level} \\
\midrule
Income & 0.586*** & 0.363*** & 0.230*** \\
& (0.019) & (0.021) & (0.019) \\
Income variability & -0.169*** & 0.032* & 0.011 \\
& (0.014) & (0.015) & (0.014) \\
Mobility & -0.207*** & -0.168*** & -0.085*** \\
& (0.013) & (0.015) & (0.013) \\
Local availability & 0.160*** & 0.226*** & 0.089*** \\
& (0.011) & (0.013) & (0.011) \\
In NYC & -0.111*** & -0.502*** & 0.501*** \\
& (0.031) & (0.035) & (0.030) \\
Median age & 0.040*** & 0.027* & 0.057*** \\
& (0.012) & (0.013) & (0.012) \\
Proportion of male & -0.022* & -0.022 & 0.010 \\
& (0.011) & (0.012) & (0.011) \\
Proportion of white & -0.191*** & -0.339*** & -0.125*** \\
& (0.016) & (0.018) & (0.015) \\
Proportion of bachelor's & -0.161*** & 0.149*** & 0.213*** \\
degree or higher & (0.017) & (0.019) & (0.016) \\
Intercept & 0.053*** & 0.184*** & -0.173*** \\
& (0.016) & (0.017) & (0.015) \\
\midrule
Observations & 6088 & 3672 & 5739 \\
R² & 0.247 & 0.401 & 0.315 \\
Adjusted R² & 0.246 & 0.399 & 0.314 \\
Residual Std. Error & 0.868 & 0.745 & 0.820 \\
F Statistic & 221.435*** & 272.347*** & 292.654*** \\
\bottomrule
\end{tabular}
\begin{flushleft}
\footnotesize
Note: * $p < 0.05$, ** $p < 0.01$, *** $p < 0.001$ (two-tailed tests).
\end{flushleft}
\end{table}

We repeat the regression analyses separately by industry, largely finding consistent results (\ref{app:appendix_c}). Income has a significant positive association with and is the strongest predictor of consumption diversity across all industries when we use entropy by brand as the measure. The same holds true using the other two measures with a few reasonable exceptions when the number of observations is limited or some cases where income and local availability have similar level of association with diversity. Overall, these results suggest that the association between income and consumption diversity is not likely to be mediated or confounded by variability, mobility, and local availability. The results provide further support for the assumptions behind the diversity hypothesis as they show that there are deeper social and cultural factors affecting consumption practices beyond simple geographic constraints.

\subsection{Replicability}

Overall, the results replicate when we use education as a measure of SES or data from Texas.

The proportion of people who have bachelor's degree or higher is significantly correlated with the three measures of diversity: the correlation coefficients are 0.133 for entropy by brand visits, 0.381 for the standard deviation of visited brands' SES, and 0.390 for entropy by brands' price level; all correlations are statistically significant at the 0.001 level (see Figure \ref{fig:fig_e4}). As shown in Table \ref{tab:tab3}, controlling for income, income variability, mobility, local availability and other demographic variables, education has a significantly positive association with diversity measured with the standard deviation of visited brands' SES and the entropy by brands' price level, but the association is negative when diversity is measured with the entropy by brand visits. To address this inconsistency, we introduce an interaction effect between income and education in our regression analyses, which comes out negative and significant (Table \ref{tab:tab_e3}). Figure \ref{fig:fig_e5} illustrates the interaction effects by showing the regression lines of income (or education) and diversity given different values of education (or income). In most cases, the associations between income or education and diversity are positive. The interaction suggests that the correlation between income (or education) and diversity is weaker for people with high education (or income). When measuring diversity with the entropy by brand visits, the correlation between education and diversity becomes negative for high income CBGs. The findings are similar if we repeat the regression analyses separately by industry (see Tables \ref{tab:tab_e4}, \ref{tab:tab_e5}, \ref{tab:tab_e6}).

All analyses conducted in Texas reveal similar patterns to those observed in New York State (\ref{app:appendix_f}). In fact, the data in Texas show stronger support for our main hypothesis. In Texas, both income and education have significant and positive correlation with all three measures of diversity, both in general and separately by industries. Notably, unlike in New York State, where the correlations between education and diversity turn negative for a few industries, the correlations remain positive in Texas. These positive associations persist even after controlling for income variability, mobility, local availability, age, gender, and race. The interaction effects between income and education vary but largely support our main hypothesis. Additionally, descriptive patterns and brand co-visit analyses in Texas mirror those observed in New York State.

In summary, the replication analyses, utilizing education as an alternative measure of SES and employing data from Texas, consistently support our primary finding that high SES is associated with diverse consumption patterns. However, these replication analyses also reveal nuanced variations within the phenomenon. Notably, in New York State, weak negative associations between education and diversity are observed for a few industries related to necessity goods, whereas in Texas, the same industries exhibit positive associations. The interaction effects of income and education on diversity are not uniformly clear, with varying directions and significance levels across states, diversity measures, and industries. These nuances, while not undermining our hypothesis, underscore the complexity of the phenomenon. It is hard to dig into these nuances with the data available for this paper. Future research with more detailed individual level data is needed to address these complexities.

\section{Discussion, Conclusion, and Limitations}

Using large-scale data of mobile tracked visits, our study reveals inequality in daily consumption: high-SES individuals consume more diversely in terms of brands and price levels than low-SES individuals. It is not true that expensive goods are for the rich and cheap goods for the poor. Rather, cheap goods are for the poor and all goods are for the rich. In other words, inconspicuous consumption, niche consumption, and brand omnivorousness are options and privileges mainly for the rich. We find that the association between diverse consumption and SES is prevalent across different industries, although stronger in industries that involve leisure and cultural expression compared to those that concern necessity goods. We further establish that the association cannot be attributed entirely to simple geographic constraints, suggesting deeper social and cultural factors.

Our contribution is both theoretical and empirical. First, combining insights from sociology, social psychology, and consumer research, we integrate the separate and to a certain degree opposing theories of cultural omnivorousness, conspicuous consumption, and inconspicuous consumption into one coherent theoretical argument. Essentially, we extend the omnivorousness argument from cultural sociology beyond cultural consumption to any consumption and we subsume omnivorousness, niche, and inconspicuous consumption under the phenomenon of diverse consumption. We argue that high-SES individuals are more omnivorous but also more niche-focused in both their cultural and material consumption because the lack of economic constraints make social and cultural aspects of expression and distinction more salient.

Second, combining data from multiple sources, we offer large-scale empirical evidence for the hypothesized links between inequality and consumption diversity. Our research used data from one US state and replicated the results for another one, which gives us confidence that the findings will qualitatively hold in other states and countries. Although our analyses are aggregate and cannot differentiate between omnivorous and niche consumption, prior research leads us to expect that both contribute to produce the observed pattern. Our findings quantify the extent to which low-income individuals are constrained in their everyday choices, and the extent to which the cornucopia and freedom of choice of market economies do not benefit all.

Nevertheless, we acknowledge several limitations to our research. One source of potential bias is the fact that we focus on brands with multiple stores and ignore a large number of unique shops and institutions. Yet, this may not be necessarily problematic. It is reasonable to assume that high-SES individuals are more likely to visit boutique shops and unique institutions such as museums and galleries, which means that we underestimate how diverse their consumption is. This means that the effects we find are even stronger in reality. Similarly, our data concern visits to brick-and-mortar shops but not online shopping. This omission may be consequential for our findings only if low-SES individuals are more likely to shop online than in person compared to high-SES individuals and/or shopping patterns differ systematically between online and in-store.

Further, we acknowledge that mobile coverage and representation, on which the mobile-tracking data depend, likely increase with higher SES. Thus, the consumption diversity we observe could be just due to the fact that we track a larger number of high SES individuals. In other words, it is not that the average high-SES individual consumes across a wider range but that we are capturing a larger number of high-SES individuals who might have just as narrow but non-overlapping consumer ranges. However, while this undermines the omnivorousness argument, it does not necessarily challenge the niche-consumption explanation.

Our data track visits to physical locations, but we do not know whether and to what extent visitors complete purchases there. It is possible that a significant proportion of the records reflect ``window shopping'' and not actual consumption. If we assume that this is more likely to be the case for low-SES individuals, this means that we are overestimating actual consumption for that group which, once again, gives us even more confidence in our findings. Certainly, it is also possible that it is high-SES individuals who engage in more leisure shopping or ``retail therapy.'' However, this tendency may be counteracted by the fact that wealthy individuals are also more likely to have shopping assistants. Unless they are live-in staff, shopping assistants will in fact boost the diversity of consumption for low-SES census block groups, and thus bias our results towards smaller, rather than larger effects. Although not available to us, SafeGraph offer data on money spent at POIs and hence, future research could replicate our analyses of visits for purchases.

Using Yelp for price comparisons entails further limitations. Price levels on Yelp are crowdsourced and hence, subjective, noisy, and potentially biased. Users' evaluations likely depend on expectations about the industry and geographical region. For instance, what can be considered expensive in Horseheads, NY may be cheap in New York City, and what is inexpensive for a new pair of jeans may be pricey for a meal. However, from our theoretical perspective centred on social distinction, relative comparisons are preferable to absolute prices and subjective evaluations more valuable than unfamiliar objective price indices. Still, high-SES areas may have more visitors and tourists contributing Yelp reviews, and their evaluations may be upwardly biased compared to residents' evaluations. Despite this, the Yelp price level estimates largely replicate the results obtained with alternative measures of consumption diversity, affirming the robustness of our findings.

Most importantly, we remind the reader that the study is conducted at the aggregate level and may not necessarily reflect individual choices and behaviour. Averaging behaviour inevitably hides much nuance and detail but may also introduce bias. We analyse census block groups and hence, we lump together 600-3000 individuals into a single unit. However, these individuals are not independent since some of them cohabit in households. Moreover, these individuals may be quite heterogeneous and the heterogeneity may be higher for census block groups with higher median income. Although we control for income variability in the regression models, the data prevent us from distinguishing the extent to which the average high-status individual is diverse in their consumption (omnivorousness) from the extent to which the high-status strata comprise diverse individuals with narrow consumption preferences (niche consumption). Research with detailed individual consumption data already provides some evidence for omnivorousness \citep{leo_correlations_2018} but we require more quantitative research of inconspicuous and niche consumption by SES. One promising direction is to extend the concepts of variety and atypicality regarding cultural preferences \citep{goldberg_what_2016} to consumption in order to quantify the relative prevalence of omnivorous and niche consumption. Another promising direction is to use qualitative or survey research to test the two supposed pathways of between- and within- social class distinction that may be influencing consumption patterns.

Overall, our research suggests that inequalities in material consumption parallel inequalities sociologists have already established regarding cultural taste: low-SES individuals are more constrained in their consumption practices than high-SES individuals. Importantly, however, the constraints are not necessarily economic but possibly cultural and social; this explains why low-SES individuals frequent expensive Gucci stores, for instance. This has an important implication. The cultural and social aspects of daily consumption could entrench economic advantages and disadvantages: the rich save by choosing to consume at the lower end, while the poor get in debt by being tempted to consume at the higher end. Consequently, to fight inequality, we require levelling policies and nudges at the cultural and social, not just economic, levels.

\section*{Abbreviations}

CBG: Census block group

NAICS: North American Industry Classification System

NYC: New York City

POI: Point of interest

SES: Socioeconomic status

\section*{Declarations}

\subsection*{Availability of data and materials}

The data that support the findings of this study are available from SafeGraph, but restrictions apply to the availability of these data, which were used under license for the current study, and so are not publicly available. Aggregate data and replication code are available in the figshare repository, \href{https://doi.org/10.6084/m9.figshare.29320598.v1}{https://doi.org/10.6084/m9.figshare.29320598.v1}

\subsection*{Competing interests}

The authors declare that they have no competing interests.

\subsection*{Funding}

The authors have no funding to declare.

\subsection*{Authors' contributions}

Y.H. and M.T. conceived the study and designed the methodology. Y.H. collected the data and performed all analyses. Y.H. and M.T. wrote and revised the manuscript.

\subsection*{Acknowledgements}

The authors are grateful to Eleanor Power, Melissa Sands, Giovanna Scarchilli, Xavier Jara, and Minsu Park for valuable detailed feedback.

\bibliographystyle{plainnat}
\bibliography{references}

\vspace{3em}

\appendix

\renewcommand{\thesection}{Appendix \Alph{section}}

\makeatletter
\renewcommand{\@seccntformat}[1]{\csname the#1\endcsname: }
\makeatother

{\Large\bfseries Appendices}

\section{Additional information about the data sample}
\label{app:appendix_a}

\setcounter{figure}{0}
\setcounter{table}{0}
\renewcommand{\thefigure}{\Alph{section}\arabic{figure}}
\renewcommand{\thetable}{\Alph{section}\arabic{table}}

\begin{table}[ht]
\centering
\caption{Four-digit NAICS industries and number of POIs included in the study.}
\label{tab:tab_a1}
\footnotesize
\begin{tabular}{lrr}
\toprule
\textbf{Four-digit industry name} & \textbf{NAICS code} & \textbf{Number of POIs} \\
\midrule
Restaurants and Other Eating Places & 7225 & 87070 \\
Personal Care Services & 8121 & 38452 \\
Other Amusement and Recreation Industries & 7139 & 15312 \\
Grocery Stores & 4451 & 15219 \\
Health and Personal Care Stores & 4461 & 13123 \\
Clothing Stores & 4481 & 12755 \\
Gasoline Stations & 4471 & 10356 \\
Museums, Historical Sites, and Similar Institutions & 7121 & 8360 \\
Drinking Places (Alcoholic Beverages) & 7224 & 6964 \\
Sporting Goods, Hobby, and Musical Instrument Stores & 4511 & 6935 \\
Jewelry, Luggage, and Leather Goods Stores & 4483 & 6906 \\
Other Miscellaneous Store Retailers & 4539 & 6901 \\
Specialty Food Stores & 4452 & 6430 \\
Beer, Wine, and Liquor Stores & 4453 & 4487 \\
Florists & 4531 & 4118 \\
Used Merchandise Stores & 4533 & 2972 \\
Office Supplies, Stationery, and Gift Stores & 4532 & 2949 \\
Consumer Goods Rental & 5322 & 2784 \\
Shoe Stores & 4482 & 2682 \\
General Merchandise Stores, including Warehouse Clubs and Supercenters & 4523 & 2503 \\
Book Stores and News Dealers & 4512 & 1946 \\
Special Food Services & 7223 & 1720 \\
Amusement Parks and Arcades & 7131 & 1272 \\
Department Stores & 4522 & 932 \\
Motion Picture and Video Industries & 5121 & 709 \\
Gambling Industries & 7132 & 491 \\
Performing Arts Companies & 7111 & 288 \\
Spectator Sports & 7112 & 190 \\
\bottomrule
\end{tabular}
\end{table}

\begin{figure}[h]
\centering
\includegraphics[width=0.6\textwidth]{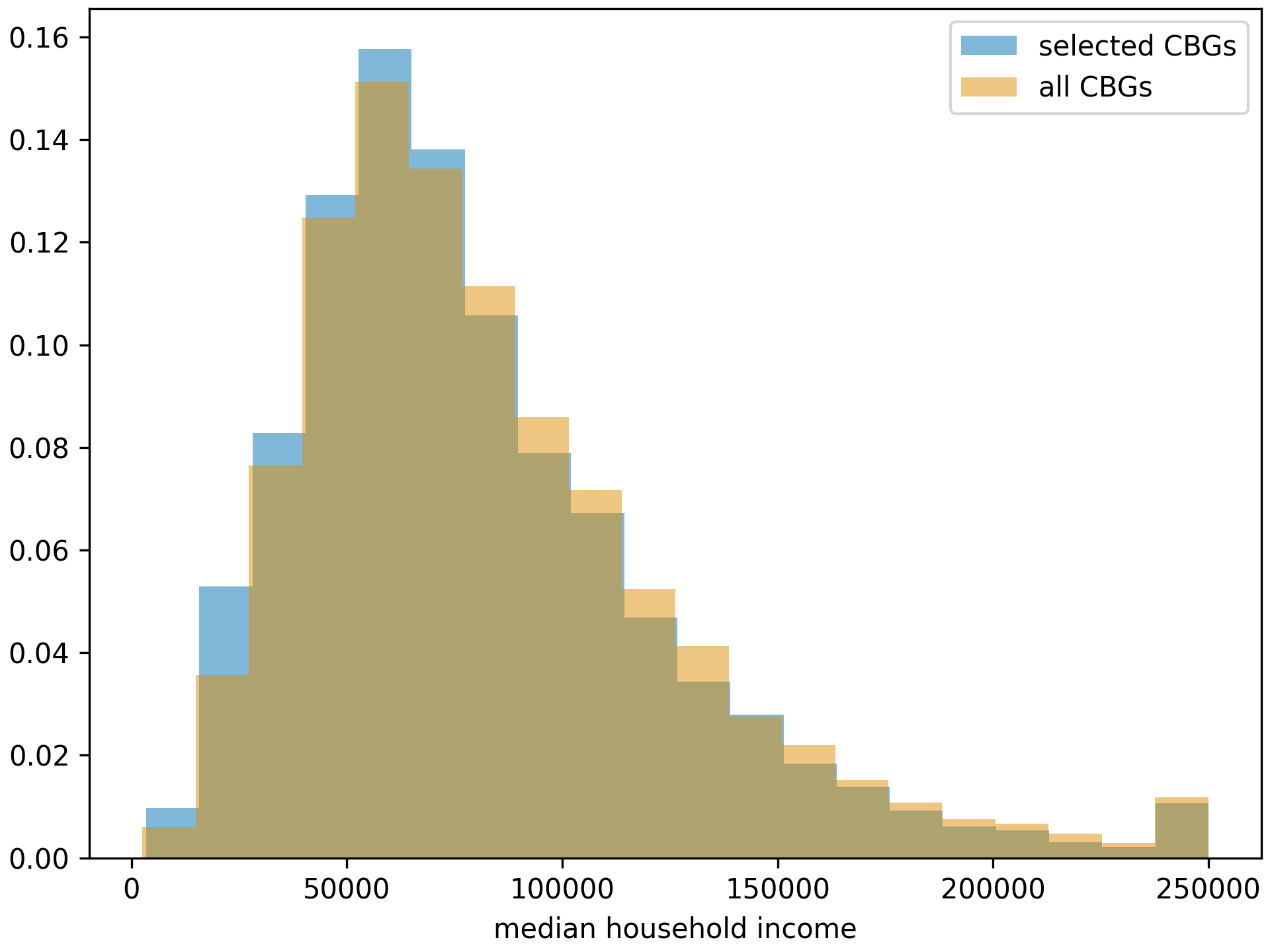}
\caption{The distribution of median household income of all 52,509 CBGs in our sample versus the selected 13,653 CBGs that have at least 100 outgoing visitors.}
\label{fig:fig_a1}
\end{figure}

As shown in Figure~\ref{fig:fig_a1}, the selected 13,653 CBGs slightly oversample lower income CBGs but still cover the full range of income levels. As our results show positive correlation between income and diversity, there is no reason to believe that the slight oversampling of lower income CBGs could significantly bias our results.

\begin{table}[h]
\centering
\caption{Correlations between median household income and six different measures of diversity in consumption.}
\label{tab:tab_a2}
\footnotesize
\begin{tabular}{lrrrrrrr}
\toprule
& \textbf{Income} & \textbf{N brands} & \textbf{Brand entropy} & \textbf{SES range} & \textbf{SES std} & \textbf{N price level} & \textbf{Price entropy} \\
\midrule
\textbf{Income} & 1 & 0.288*** & 0.292*** & 0.365*** & 0.471*** & 0.296*** & 0.358*** \\
\textbf{N brands} & 0.288*** & 1 & 0.918*** & 0.507*** & 0.321*** & 0.459*** & 0.262*** \\
\textbf{Brand entropy} & 0.292*** & 0.918*** & 1 & 0.543*** & 0.398*** & 0.493*** & 0.369*** \\
\textbf{SES range} & 0.365*** & 0.507*** & 0.543*** & 1 & 0.809*** & 0.398*** & 0.409*** \\
\textbf{SES std} & 0.471*** & 0.321*** & 0.398*** & 0.809*** & 1 & 0.363*** & 0.466*** \\
\textbf{N price level} & 0.296*** & 0.459*** & 0.493*** & 0.398*** & 0.363*** & 1 & 0.742*** \\
\textbf{Price entropy} & 0.358*** & 0.262*** & 0.369*** & 0.409*** & 0.466*** & 0.742*** & 1 \\
\bottomrule
\end{tabular}
\begin{flushleft}
\footnotesize
Note: *** $p < 0.001$ (two-tailed tests).
\end{flushleft}
\end{table}

We tried six measures of diversity, including the number of brands visited (\emph{N brands} in Table~\ref{tab:tab_a2}), the normalized Shannon entropy by brands (\emph{Brand entropy}), the range of brands' SES (\emph{SES range}), the standard deviation of brands' SES (\emph{SES std}), the number of price levels visited (\emph{N price level}), and normalized Shannon entropy by price levels (\emph{Price entropy}). As Table~\ref{tab:tab_a2} shows, all measures are significantly correlated with each other and the median income of the CBG. Among the most correlated pairs (N brands \& Brand entropy, SES range \& SES std, and N price level \& Price entropy), we chose the ones that represent more information to report and use in further analyses.

\section{Predicting CBGs' median household income from consumption patterns}
\label{app:appendix_b}

\setcounter{figure}{0}
\setcounter{table}{0}
\renewcommand{\thefigure}{\Alph{section}\arabic{figure}}
\renewcommand{\thetable}{\Alph{section}\arabic{table}}

We use a LASSO regression model to predict the median household income of the CBGs with the proportion of outgoing visitors of the CBGs to the 924 brands. Following the standard practice, we keep 20 percent of the observations as a test set to examine the out-of-sample prediction accuracy. The LASSO regression model is an extension of the linear regression model estimated with ordinary least squares (OLS) that reduces overfitting and variance and increases prediction accuracy and interpretability. It adds a penalty term to the loss function of OLS regression which shrinks some of the coefficients towards zero. Let us take as an example the multiple linear regression:

\begin{equation*}
y = \beta_0 + \beta_1 x_1 + \beta_2 x_2 + \cdots + \beta_p x_p + \varepsilon
\end{equation*}

where $y$ represents the outcome variable, $\beta_0$ is the intercept, $x_j$ represents the $j$th predictor with coefficient $\beta_j$, $\varepsilon$ represents the error term, and $p$ -- the number of predictors. The OLS regression model aims to minimize the residual sum of squares (RSS):

\begin{equation*}
RSS = \sum_{i=1}^{n}(y_i - \hat{y}_i)^2 = \sum_{i=1}^{n}\left(y_i - \beta_0 - \sum_{j=1}^{p}\beta_j x_{ij}\right)^2
\end{equation*}

where $y_i$ represents the actual value of the $i$th observation of $y$, $\hat{y}_i$ represents the predicted value, and $n$ represents the number of observations. The LASSO model, however, aims to minimize the following function:

\begin{equation*}
RSS + \lambda\sum_{j=1}^{p}|\beta_j|
\end{equation*}

where $\lambda \geq 0$ is a tuning parameter to be determined separately. In this paper, we use the standard practice of using five-fold cross validation to select the value of $\lambda$. Due to the penalty term $\lambda\sum_{j=1}^{p}|\beta_j|$, to minimise the loss function, some coefficients need to shrink, sometimes all the way to zero. The shrinkage prevents overfitting and performs variable selection.

\begin{figure}[h]
\centering
\includegraphics[width=0.5\textwidth]{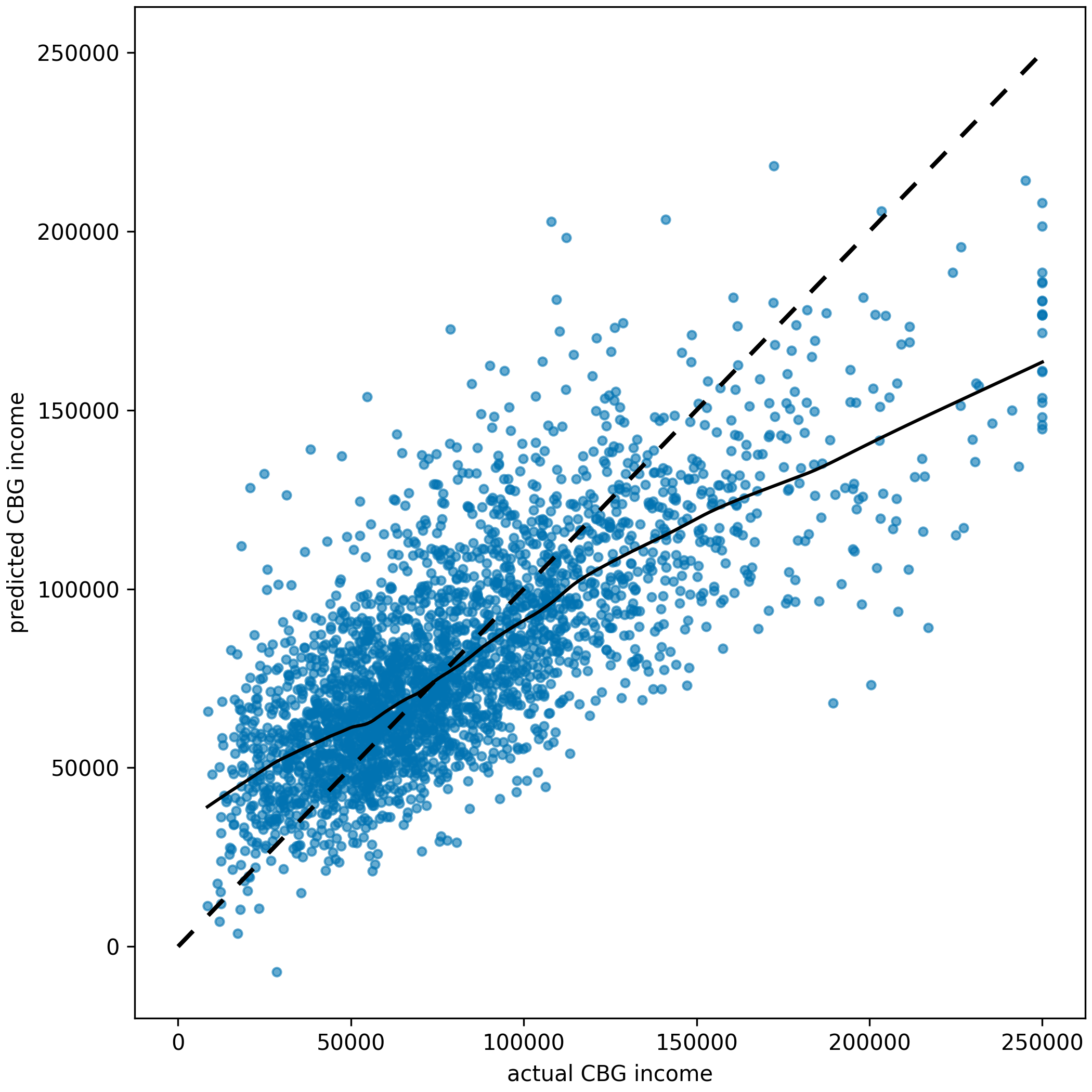}
\caption{Correlation between predicted and actual income in the test set. The solid line shows the LOESS fit and the dashed line -- the ideal 1:1 relation.}
\label{fig:fig_b1}
\end{figure}

The LASSO regression model that predicts CBGs' median household income based on the proportion of outgoing visitors of the CBGs to the 924 brands has strong predictive power. For the training sample, the model explains 0.590 of the variances. Using the 355 brands selected by the model, the out-of-sample correlation between the predicted income and actual income is 0.748 ($p < 0.001$). Figure~\ref{fig:fig_b1} shows the correlation between predicted and actual income in the test set. Such strong prediction performance suggests that SES is strongly associated with consumption preferences, even if consumption preferences are not determined by the economic constraints driven by the product prices. Nevertheless, the figure also shows that the predictions perform notably worse for CBGs with the highest income. One possible explanation for this is that the wealthiest individuals seek elite and authentic consumption experience and thus avoid brands.

\clearpage

\section{Results from separate regression analyses by industry}
\label{app:appendix_c}

\setcounter{figure}{0}
\setcounter{table}{0}
\renewcommand{\thefigure}{\Alph{section}\arabic{figure}}
\renewcommand{\thetable}{\Alph{section}\arabic{table}}

\begin{table}[ht]
\centering
\caption{Entropy by brand}
\label{tab:tab_c1}
\tiny
\setlength{\tabcolsep}{3pt}
\begin{tabular}{lrrrrrrrrr}
\toprule
& \textbf{Amusement,} & \textbf{Clothing and} & \textbf{Food Services} & \textbf{Food and} & \textbf{Gasoline} & \textbf{General} & \textbf{Health and} & \textbf{Miscellaneous} & \textbf{Sporting} \\
& \textbf{Gambling,} & \textbf{Clothing} & \textbf{and Drinking} & \textbf{Beverage} & \textbf{Stations} & \textbf{Merchandise} & \textbf{Personal Care} & \textbf{Store} & \textbf{Goods,} \\
& \textbf{and Recreation} & \textbf{Accessories} & \textbf{Places} & \textbf{Stores} & & \textbf{Stores} & \textbf{Stores} & \textbf{Retailers} & \textbf{Hobby} \\
& \textbf{Industries} & \textbf{Stores} & & & & & & & \\
\midrule
Income & 0.406*** & 0.376*** & 0.440*** & 0.347*** & 0.437*** & 0.340*** & 0.260*** & 0.408*** & 0.289*** \\
& (0.045) & (0.060) & (0.025) & (0.035) & (0.026) & (0.041) & (0.040) & (0.055) & (0.077) \\
Income & -0.120*** & -0.136** & -0.115*** & -0.106*** & -0.080*** & -0.051 & -0.090** & -0.146*** & -0.254*** \\
variability & (0.034) & (0.042) & (0.019) & (0.025) & (0.021) & (0.034) & (0.029) & (0.039) & (0.061) \\
Mobility & 0.013 & -0.074 & -0.106*** & 0.040 & 0.123*** & -0.027 & 0.171*** & 0.101 & 0.012 \\
& (0.052) & (0.043) & (0.020) & (0.022) & (0.024) & (0.027) & (0.034) & (0.058) & (0.069) \\
Local & 0.135*** & 0.205*** & 0.186*** & 0.028 & 0.120*** & 0.073*** & 0.152*** & 0.122*** & 0.127* \\
availability & (0.031) & (0.041) & (0.015) & (0.020) & (0.015) & (0.020) & (0.024) & (0.036) & (0.049) \\
In NYC & 0.224** & 0.437*** & 0.118** & -0.159** & -0.387*** & -0.508*** & 0.329*** & -0.597*** & -0.204 \\
& (0.079) & (0.108) & (0.042) & (0.057) & (0.049) & (0.071) & (0.064) & (0.099) & (0.132) \\
Median age & 0.015 & 0.066 & 0.005 & 0.004 & 0.002 & 0.033 & -0.045 & 0.023 & 0.157** \\
& (0.031) & (0.042) & (0.016) & (0.021) & (0.017) & (0.021) & (0.025) & (0.037) & (0.048) \\
Proportion of & -0.005 & 0.022 & -0.020 & -0.026 & -0.030 & -0.034 & -0.046 & 0.014 & 0.083 \\
male & (0.029) & (0.040) & (0.015) & (0.020) & (0.016) & (0.020) & (0.025) & (0.035) & (0.045) \\
Proportion of & -0.087 & -0.140* & -0.174*** & -0.150*** & -0.126*** & -0.093** & -0.034 & -0.067 & -0.178* \\
white & (0.046) & (0.063) & (0.021) & (0.028) & (0.022) & (0.031) & (0.034) & (0.052) & (0.074) \\
Proportion of & 0.080 & -0.002 & -0.092*** & -0.157*** & -0.247*** & -0.077* & -0.076* & -0.057 & 0.097 \\
bachelor's degree & (0.042) & (0.060) & (0.022) & (0.031) & (0.025) & (0.034) & (0.035) & (0.052) & (0.069) \\
or higher & & & & & & & & & \\
Intercept & 0.163*** & 0.075 & 0.083*** & 0.150*** & 0.344*** & 0.282*** & -0.013 & 0.500*** & 0.373*** \\
& (0.042) & (0.063) & (0.021) & (0.028) & (0.019) & (0.029) & (0.033) & (0.045) & (0.065) \\
\midrule
Observations & 814 & 523 & 3192 & 2071 & 2559 & 1217 & 1578 & 608 & 362 \\
R² & 0.264 & 0.215 & 0.191 & 0.058 & 0.162 & 0.123 & 0.097 & 0.187 & 0.138 \\
Adjusted R² & 0.256 & 0.202 & 0.188 & 0.054 & 0.159 & 0.116 & 0.092 & 0.175 & 0.116 \\
Residual Std. & 0.828 & 0.926 & 0.857 & 0.894 & 0.763 & 0.698 & 0.943 & 0.858 & 0.916 \\
Error & & & & & & & & & \\
F Statistic & 32.075*** & 15.656*** & 83.301*** & 14.218*** & 54.602*** & 18.749*** & 18.742*** & 15.279*** & 6.261*** \\
\bottomrule
\end{tabular}
\begin{flushleft}
\footnotesize
Note: *** $p < 0.001$, ** $p < 0.01$, * $p < 0.05$ (two-tailed tests). No regression model for the industry \emph{Motion Picture and Video Industries} due to limited number of observations with price and SES data.
\end{flushleft}
\end{table}

\begin{table}[h]
\centering
\caption{Standard deviation of brands' SES}
\label{tab:tab_c2}
\tiny
\setlength{\tabcolsep}{3pt}
\begin{tabular}{lrrrrrrrrr}
\toprule
& \textbf{Amusement,} & \textbf{Clothing and} & \textbf{Food Services} & \textbf{Food and} & \textbf{Gasoline} & \textbf{General} & \textbf{Health and} & \textbf{Miscellaneous} & \textbf{Sporting} \\
& \textbf{Gambling,} & \textbf{Clothing} & \textbf{and Drinking} & \textbf{Beverage} & \textbf{Stations} & \textbf{Merchandise} & \textbf{Personal Care} & \textbf{Store} & \textbf{Goods,} \\
& \textbf{and Recreation} & \textbf{Accessories} & \textbf{Places} & \textbf{Stores} & & \textbf{Stores} & \textbf{Stores} & \textbf{Retailers} & \textbf{Hobby} \\
& \textbf{Industries} & \textbf{Stores} & & & & & & & \\
\midrule
Income & 0.078 & 0.190* & 0.272*** & 0.276*** & 0.006 & 0.218** & 0.275*** & 0.231* & 0.429** \\
& (0.076) & (0.081) & (0.032) & (0.067) & (0.054) & (0.072) & (0.075) & (0.096) & (0.127) \\
Income & 0.069 & 0.007 & 0.060* & -0.032 & -0.061 & 0.112 & -0.089 & 0.152* & -0.102 \\
variability & (0.061) & (0.057) & (0.025) & (0.048) & (0.044) & (0.061) & (0.055) & (0.068) & (0.083) \\
Mobility & 0.272* & -0.050 & -0.032 & 0.092 & 0.034 & 0.030 & 0.211** & 0.053 & 0.388** \\
& (0.130) & (0.069) & (0.027) & (0.065) & (0.056) & (0.059) & (0.076) & (0.125) & (0.123) \\
Local & 0.145* & 0.131* & 0.288*** & 0.264*** & 0.366*** & 0.091* & 0.275*** & 0.176** & 0.357*** \\
availability & (0.058) & (0.058) & (0.021) & (0.040) & (0.031) & (0.039) & (0.048) & (0.064) & (0.076) \\
In NYC & 0.141 & 0.154 & -0.389*** & -0.223 & -0.293** & -0.030 & 0.050 & -0.319 & -0.226 \\
& (0.144) & (0.145) & (0.055) & (0.144) & (0.105) & (0.150) & (0.128) & (0.218) & (0.232) \\
Median age & -0.027 & 0.008 & 0.066** & 0.038 & 0.026 & 0.068 & -0.038 & 0.016 & 0.142* \\
& (0.058) & (0.054) & (0.021) & (0.038) & (0.036) & (0.038) & (0.046) & (0.062) & (0.070) \\
Proportion of & 0.062 & -0.118* & -0.019 & 0.047 & 0.005 & -0.007 & -0.026 & 0.083 & 0.109 \\
male & (0.054) & (0.052) & (0.019) & (0.046) & (0.034) & (0.043) & (0.045) & (0.067) & (0.070) \\
Proportion of & -0.088 & -0.258** & -0.226*** & -0.456*** & -0.124** & -0.262*** & 0.003 & -0.126 & -0.194 \\
white & (0.095) & (0.082) & (0.028) & (0.069) & (0.046) & (0.067) & (0.074) & (0.105) & (0.111) \\
Proportion of & 0.165 & -0.016 & 0.070* & -0.022 & 0.055 & 0.008 & -0.092 & -0.027 & -0.063 \\
bachelor's degree & (0.085) & (0.080) & (0.028) & (0.063) & (0.048) & (0.060) & (0.069) & (0.099) & (0.108) \\
or higher & & & & & & & & & \\
Intercept & 0.177* & 0.156 & 0.265*** & -0.018 & 0.331*** & 0.075 & 0.157* & 0.369*** & 0.381*** \\
& (0.088) & (0.084) & (0.028) & (0.058) & (0.039) & (0.055) & (0.071) & (0.084) & (0.098) \\
\midrule
Observations & 150 & 246 & 1651 & 295 & 684 & 281 & 385 & 147 & 98 \\
R² & 0.278 & 0.156 & 0.284 & 0.365 & 0.186 & 0.208 & 0.145 & 0.237 & 0.363 \\
Adjusted R² & 0.231 & 0.124 & 0.280 & 0.345 & 0.175 & 0.181 & 0.124 & 0.187 & 0.297 \\
Residual Std. & 0.657 & 0.842 & 0.802 & 0.651 & 0.804 & 0.641 & 0.892 & 0.727 & 0.706 \\
Error & & & & & & & & & \\
F Statistic & 5.985*** & 4.863*** & 72.245*** & 18.226*** & 17.063*** & 7.892*** & 7.056*** & 4.727*** & 5.562*** \\
\bottomrule
\end{tabular}
\begin{flushleft}
\footnotesize
Note: *** $p < 0.001$, ** $p < 0.01$, * $p < 0.05$ (two-tailed tests). No regression model for the industry \emph{Motion Picture and Video Industries} due to limited number of observations with price and SES data.
\end{flushleft}
\end{table}

\begin{table}[ht]
\centering
\caption{Entropy by brands' price level}
\label{tab:tab_c3}
\tiny
\setlength{\tabcolsep}{3pt}
\begin{tabular}{lrrrrrrrrr}
\toprule
& \textbf{Amusement,} & \textbf{Clothing and} & \textbf{Food Services} & \textbf{Food and} & \textbf{Gasoline} & \textbf{General} & \textbf{Health and} & \textbf{Miscellaneous} & \textbf{Sporting} \\
& \textbf{Gambling,} & \textbf{Clothing} & \textbf{and Drinking} & \textbf{Beverage} & \textbf{Stations} & \textbf{Merchandise} & \textbf{Personal Care} & \textbf{Store} & \textbf{Goods,} \\
& \textbf{and Recreation} & \textbf{Accessories} & \textbf{Places} & \textbf{Stores} & & \textbf{Stores} & \textbf{Stores} & \textbf{Retailers} & \textbf{Hobby} \\
& \textbf{Industries} & \textbf{Stores} & & & & & & & \\
\midrule
Income & 0.299* & 0.099 & 0.197*** & 0.215*** & 0.346*** & 0.386*** & -0.054 & 0.106 & 0.035 \\
& (0.137) & (0.059) & (0.024) & (0.034) & (0.034) & (0.043) & (0.043) & (0.062) & (0.088) \\
Income & -0.155 & -0.004 & 0.002 & -0.066** & -0.024 & -0.012 & -0.015 & -0.101* & -0.130 \\
variability & (0.121) & (0.042) & (0.018) & (0.025) & (0.027) & (0.035) & (0.031) & (0.044) & (0.070) \\
Mobility & -0.018 & -0.003 & 0.018 & -0.005 & 0.006 & 0.004 & 0.131*** & -0.057 & 0.009 \\
& (0.142) & (0.043) & (0.020) & (0.022) & (0.030) & (0.028) & (0.037) & (0.064) & (0.075) \\
Local & 0.166 & 0.163*** & 0.129*** & 0.079*** & 0.179*** & 0.110*** & 0.168*** & 0.161*** & 0.172** \\
availability & (0.100) & (0.041) & (0.015) & (0.020) & (0.019) & (0.021) & (0.027) & (0.040) & (0.056) \\
In NYC & -0.114 & 0.511*** & 0.179*** & -0.452*** & 0.297*** & 0.161* & 0.388*** & -0.296** & 0.118 \\
& (0.261) & (0.106) & (0.041) & (0.056) & (0.062) & (0.075) & (0.069) & (0.108) & (0.147) \\
Median age & -0.077 & 0.083* & 0.045** & -0.038 & 0.011 & 0.078*** & -0.012 & 0.005 & 0.120* \\
& (0.091) & (0.042) & (0.016) & (0.021) & (0.022) & (0.022) & (0.027) & (0.041) & (0.057) \\
Proportion of & -0.056 & 0.033 & 0.024 & -0.006 & 0.007 & 0.014 & 0.078** & -0.034 & 0.013 \\
male & (0.091) & (0.040) & (0.014) & (0.020) & (0.021) & (0.022) & (0.027) & (0.039) & (0.052) \\
Proportion of & 0.074 & -0.025 & -0.029 & 0.066* & -0.168*** & -0.100** & 0.045 & -0.043 & -0.256** \\
white & (0.142) & (0.064) & (0.021) & (0.028) & (0.028) & (0.032) & (0.037) & (0.058) & (0.084) \\
Proportion of & 0.050 & 0.197** & 0.273*** & -0.037 & -0.210*** & 0.125*** & 0.204*** & -0.078 & 0.220** \\
bachelor's degree & (0.127) & (0.060) & (0.022) & (0.031) & (0.031) & (0.036) & (0.038) & (0.058) & (0.079) \\
or higher & & & & & & & & & \\
Intercept & 0.237 & -0.080 & -0.030 & 0.254*** & -0.009 & 0.015 & -0.121*** & 0.263*** & 0.043 \\
& (0.124) & (0.063) & (0.021) & (0.027) & (0.024) & (0.030) & (0.036) & (0.051) & (0.073) \\
\midrule
Observations & 178 & 512 & 3191 & 1980 & 2098 & 1195 & 1442 & 539 & 330 \\
R² & 0.090 & 0.232 & 0.257 & 0.119 & 0.151 & 0.240 & 0.118 & 0.078 & 0.122 \\
Adjusted R² & 0.041 & 0.218 & 0.255 & 0.115 & 0.147 & 0.234 & 0.113 & 0.062 & 0.098 \\
Residual Std. & 1.113 & 0.913 & 0.833 & 0.875 & 0.872 & 0.727 & 0.975 & 0.912 & 0.981 \\
Error & & & & & & & & & \\
F Statistic & 1.839 & 16.828*** & 122.259*** & 29.532*** & 41.309*** & 41.569*** & 21.330*** & 4.982*** & 4.960*** \\
\bottomrule
\end{tabular}
\begin{flushleft}
\footnotesize
Note: *** $p < 0.001$, ** $p < 0.01$, * $p < 0.05$ (two-tailed tests). No regression model for the industry \emph{Motion Picture and Video Industries} due to limited number of observations with price and SES data.
\end{flushleft}
\end{table}

\clearpage

\section{Testing for niche consumption patterns}
\label{app:appendix_d}

\setcounter{figure}{0}
\setcounter{table}{0}
\renewcommand{\thefigure}{\Alph{section}\arabic{figure}}
\renewcommand{\thetable}{\Alph{section}\arabic{table}}

\subsection{Network Embedding}

To test for niche consumption patterns, we use data on brand co-visits. We normalize the data such that we obtain a 924×924 matrix where each $i,j$ entry is the proportion of visitors from brand $i$ who also visited brand $j$ within that month.

To analyse the data, we first use the graph embedding method node2vec \citep{grover_node2vec_2016, goyal_graph_2018}. This is a semi-supervised method that learns a mapping of nodes (the brands, in this case) to a lower-dimensional vector space. Distances in this new embedding space reflect neighbourhood similarity in the original network. The method uses a biased random-walk procedure to explore each node's neighbourhood. The procedure relies on two parameters, $p$ and $q$, which control the extent to which the neighbourhood is defined locally, in the sense of a proximate community, or globally, in the sense of the network concept of structural equivalence. We set low $p = 1$ and larger $q = 2$ in order to map nodes by proximity in the network -- the more relevant concept for the definition of niche consumption.

After compressing the matrix to 128 dimensions, we visualize the results with t-SNE (t-distributed stochastic neighbour embedding) in two dimensions. This method depicts nodes that are similar in the 128 dimensions as points that are closer together on the plot than nodes that are dissimilar. Additionally, we identify clusters by applying $k$-means clustering on the 128-dimension embedding. This method groups the nodes into $k$ clusters by assigning each node to the cluster whose current centroid is closest. We attempted to use the elbow method to determine the number of clusters $k$ but since no clear elbow transpired, we chose the clustering that mapped best onto the t-SNE visualization.

We note that we explored various parameters, data processing, and methods. Specifically, we tested different combination of parameters for node2vec: number of dimensions = (32, 128, 256); $p$ = (0.25, 0.5, 1), and $q$ = (1, 2, 64). We also tested different python implementations of node2vec, as well as additional algorithms: the Walktrap community detection algorithm \citep{pons_computing_2006} and structural deep network embedding, or SDNE \citep{wang_structural_2016}. Finally, we also ran these analyses on differently trimmed network, with minimum threshold of ties for 1, 5, and 10 co-visits. We do not report the results from these additional analyses here as they are not notably different.

\subsection{Results}

Aggregate data does not allow us to test for omnivorousness. Nevertheless, prior research using individual-level mobility and banking transaction data offers supporting evidence with regards to product and service categories \citep{leo_correlations_2018} and merchants \citep{dong_purchase_2020}. Although individual-level data are also preferable to test for niche consumption, we can still use the aggregate data to test for expected macro-level patterns. Namely, if niche consumption is more common for high-SES individuals, then we will observe more numerous, smaller, and more clearly defined consumption communities for those from high-SES CBGs; these communities will also vary greatly in average price level. In contrast, those from low-SES CBGs will be visiting a larger grouping of brands at low price level.

\begin{figure}[h]
\centering
\includegraphics[width=\textwidth]{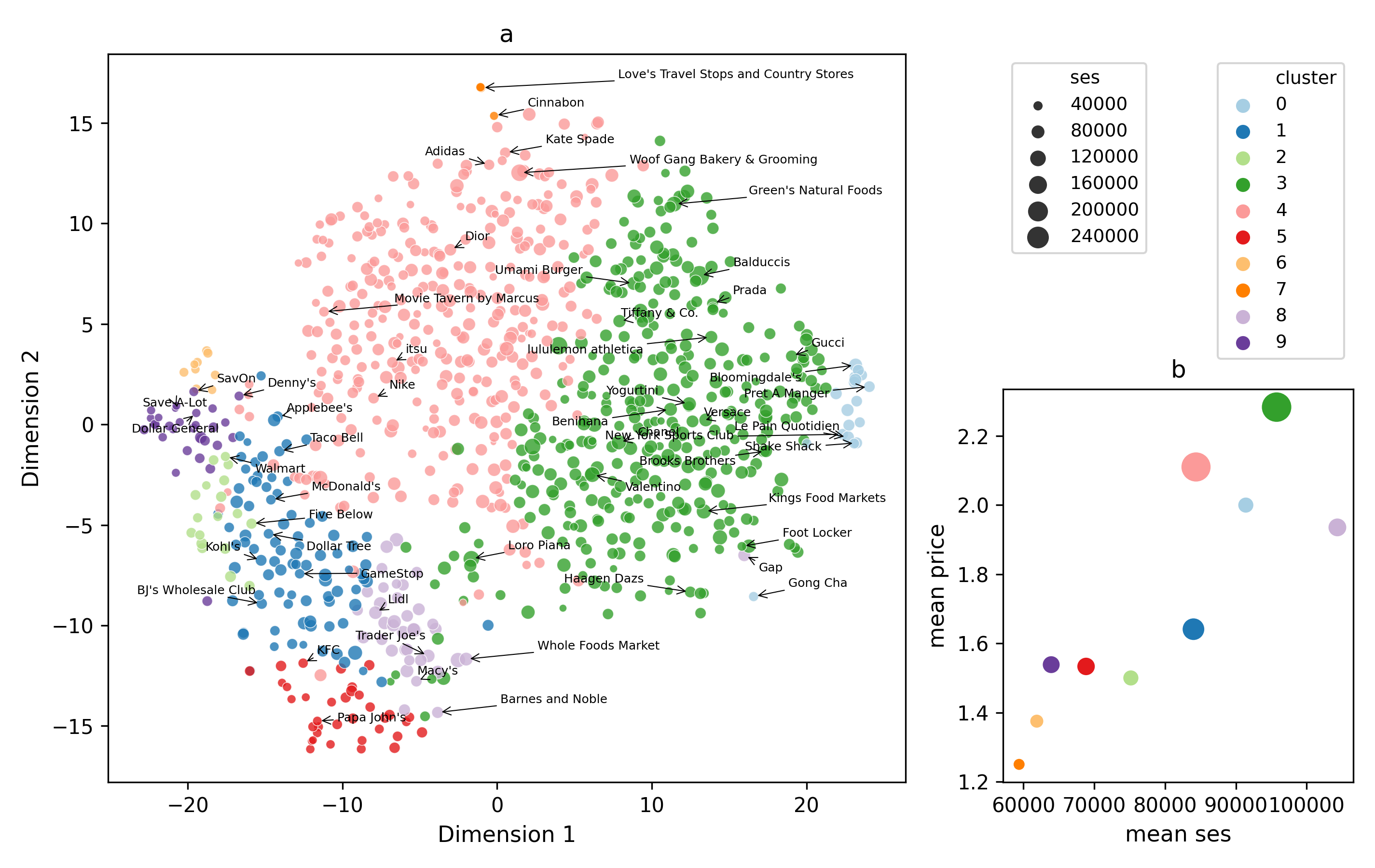}
\caption{Niche consumption analysis: a) t-SNE visualization and $k$-means clustering of the brand co-visit network after node2vec embedding to 128 dimensions; b) mean SES and price for the brands in each cluster. In panel a, the marker size represents the brand's SES, as the legend indicates. In panel b, the marker size is proportional to the square root of the number of brands in each cluster.}
\label{fig:fig_d1}
\end{figure}

The results of the network embedding analysis are presented in Figure~\ref{fig:fig_d1}. Apart from a couple of exceptions (e.g., the Pret A Manger/Bloomingdales cluster), the t-SNE plot does not identify distinct consumption niches. Additionally, mapping the 10 clusters identified by the $k$-means algorithm by mean SES and price does not reveal the expected pattern of niche consumption by SES. Instead of fewer and larger clusters at low SES, we observe the opposite -- large consumer clusters for middle and upper-middle SES and more numerous and smaller clusters for low SES. The high-SES clusters display more price-level dispersion but the patterns are not that clear. Although we don't observe the aggregate co-visit patterns we expect from niche consumption, the definite test for it should be done with individual-level data.

\section{Replicating the analyses with education}
\label{app:appendix_e}

\setcounter{figure}{0}
\setcounter{table}{0}
\renewcommand{\thefigure}{\Alph{section}\arabic{figure}}
\renewcommand{\thetable}{\Alph{section}\arabic{table}}

\begin{figure}[hp]
\centering
\includegraphics[width=0.45\textwidth]{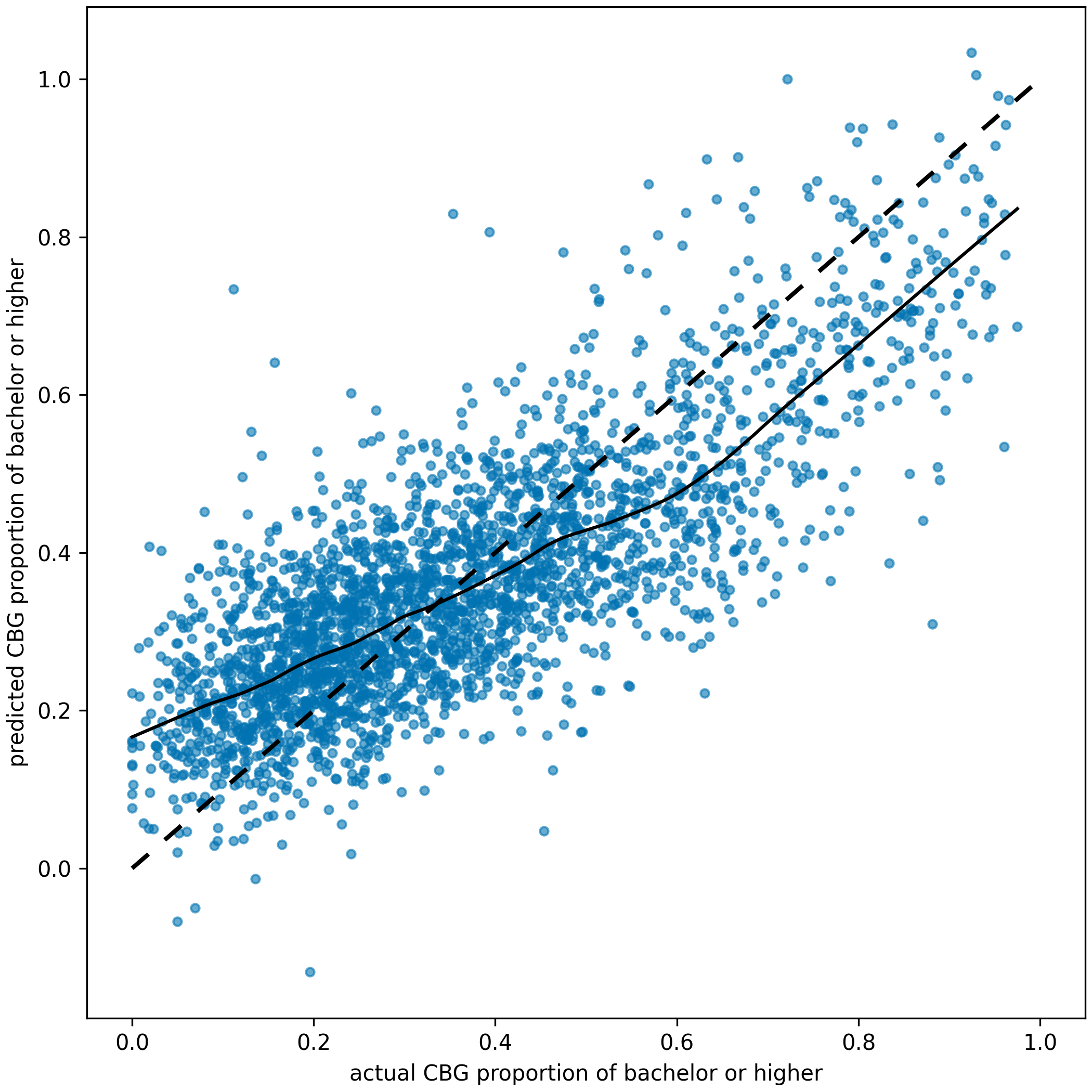}
\caption{Correlation between predicted and actual education in the test set. The solid line shows the LOESS fit and the dashed line -- the ideal 1:1 relation. Correlation = 0.792 ($p < 0.001$); variance explained = 0.655.}
\label{fig:fig_e1}
\end{figure}

\begin{figure}[h]
\centering
\includegraphics[width=\textwidth]{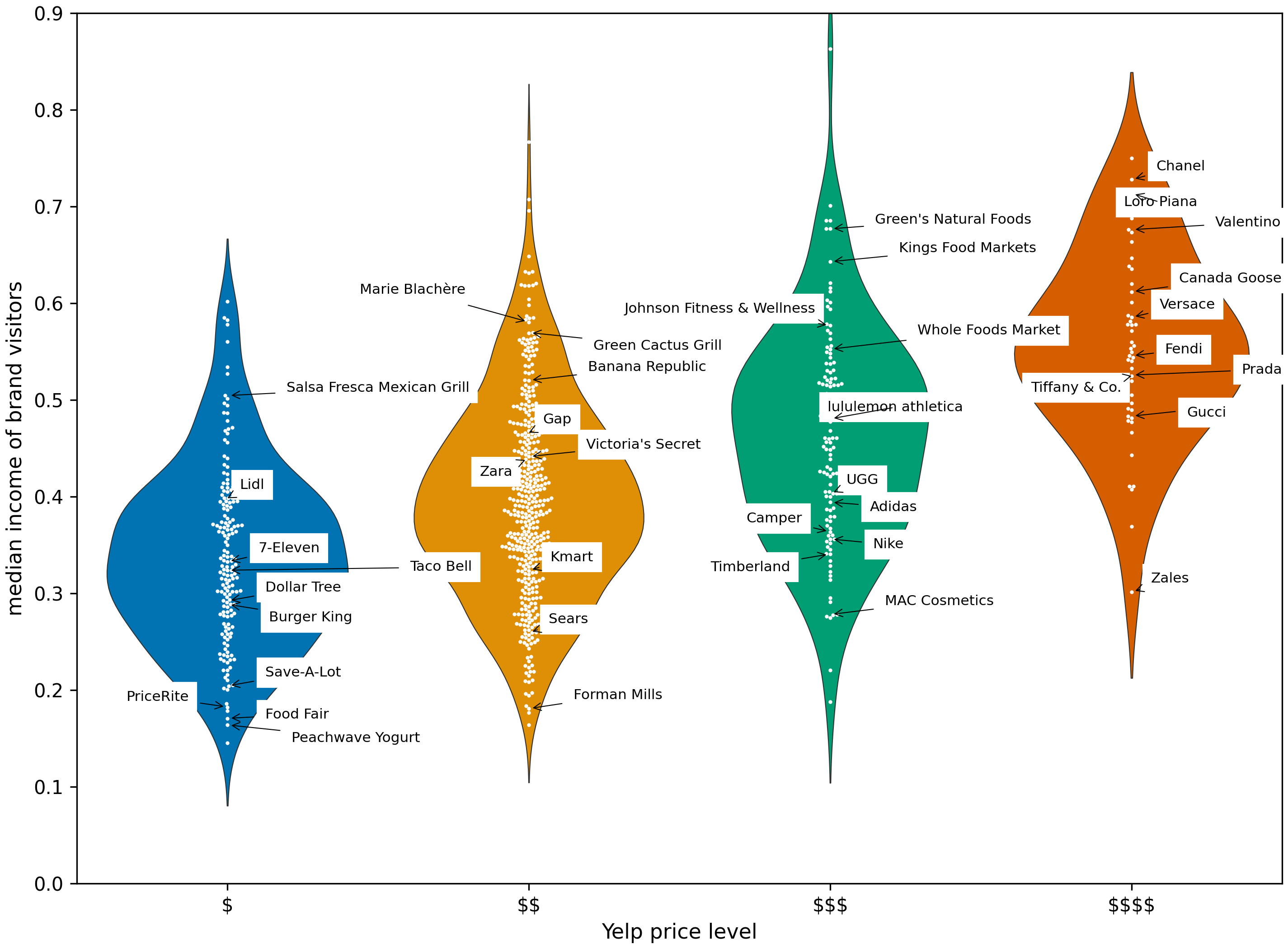}
\caption{Relation between brands' Yelp price level and median proportion of brand visitors with bachelor's or higher degree.}
\label{fig:fig_e2}
\end{figure}

\begin{figure}[hp]
\centering
\includegraphics[width=\textwidth]{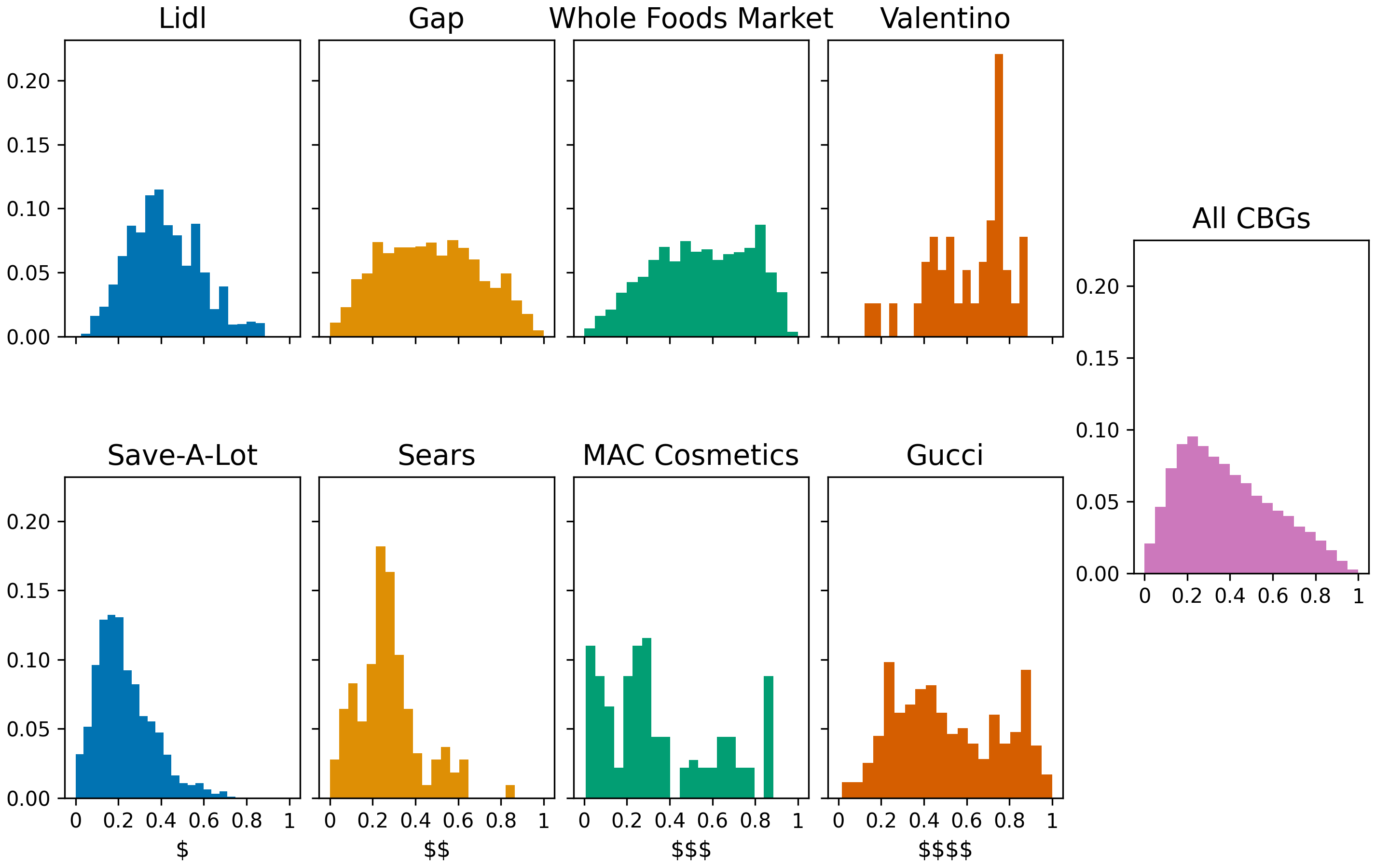}
\caption{Distribution of proportion of brand visitors with bachelor's or higher degree for several typical brands and all CBGs in our sample.}
\label{fig:fig_e3}
\end{figure}

For the CBGs in our sample, the association between median household income and proportion of people with bachelor's or higher degree is 0.710 ($p < 0.001$). Like income, using a simple LASSO regression model, we can predict proportion of people who have bachelor's or higher degree of the CBGs with the proportion of outgoing visitors of the CBGs to the brands, obtaining an out-of-sample correlation between predicted and actual proportion of people who have bachelor's degree or higher of 0.792 ($p < 0.001$, Figure~\ref{fig:fig_e1}). The distribution of the median proportion of brand visitors with bachelor's or higher degree for different Yelp price levels and for some typical brands are similar with income (Figures~\ref{fig:fig_e2} and~\ref{fig:fig_e3}).

Table \ref{tab:tab_e1} shows the association between CBGs' proportion of brand visitors with bachelor's or higher degree and the diversity measures by industry. The results largely replicate the findings with income: there are significant associations between education and the diversity measures in most of the industries. For three industries concerning necessity goods (\emph{Food and Beverage Stores, General Merchandise Stores, Gasoline Stations}) the associations between education and some measures of diversity are negative but relatively weak. Table \ref{tab:tab_e2} shows the associations between CBGs' proportion of brand visitors with bachelor's or higher degree and diversity in consumption by CBGs in and outside New York City, showing the same trends as income.

\begin{figure}[h]
\centering
\includegraphics[width=\textwidth]{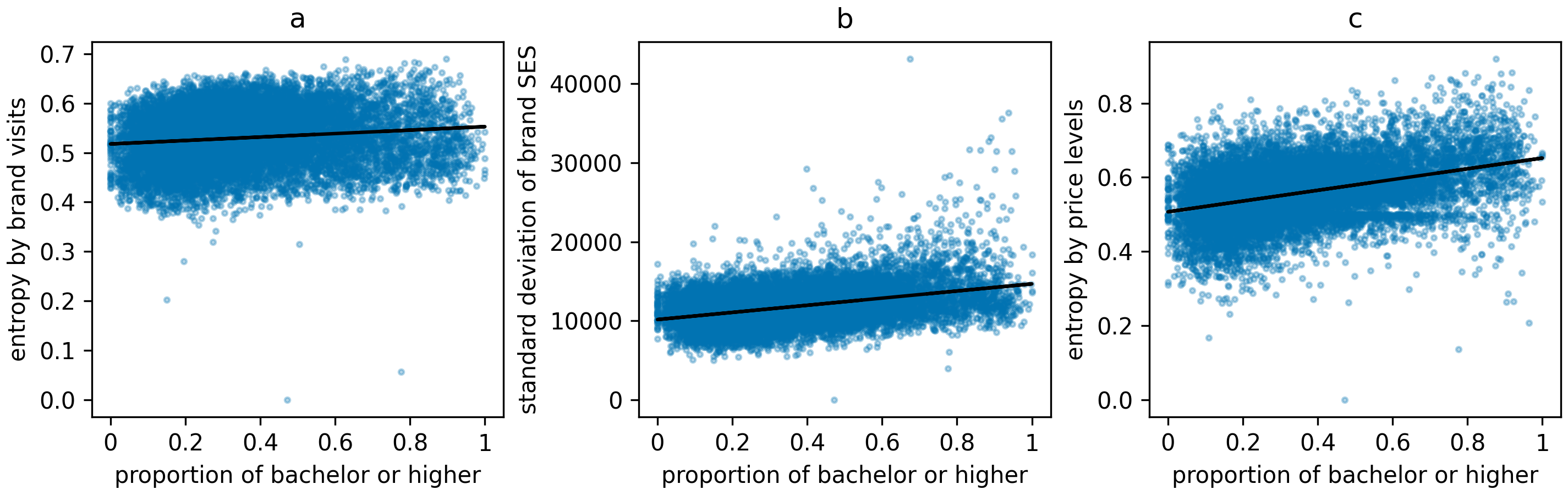}
\caption{Correlation between CBGs' proportion of brand visitors with bachelor's or higher degree and three measures of consumption diversity.}
\label{fig:fig_e4}
\end{figure}

\begin{figure}[h]
\centering
\includegraphics[width=\textwidth]{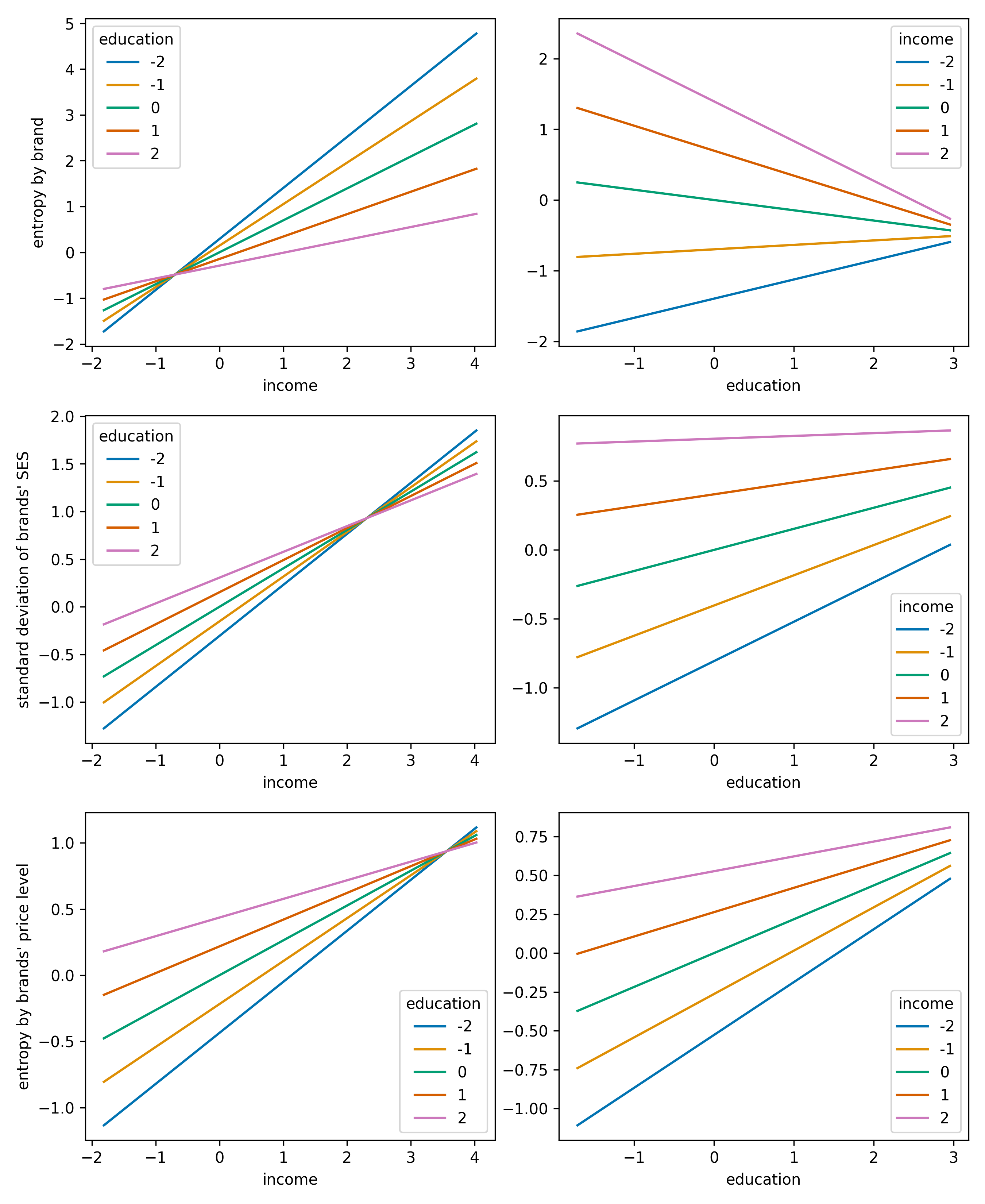}
\caption{The interaction effects between income and education on consumption diversity. The values are standardised. For example, -2 means two standard deviations lower than the mean value. The range of income and education on the x-axis are the actual ranges from the data.}
\label{fig:fig_e5}
\end{figure}

\begin{table}
\centering
\caption{The associations between CBGs' proportion of residents with bachelor's or higher degree and diversity in consumption by industry.}
\label{tab:tab_e1}
\footnotesize
\setlength{\tabcolsep}{4pt}
\begin{tabular}{lrrrrrr}
\toprule
\textbf{Industry} & \multicolumn{3}{c}{\textbf{Associations with education}} & \multicolumn{3}{c}{\textbf{Industry characteristics}} \\
\cmidrule(lr){2-4} \cmidrule(lr){5-7}
& \textbf{Entropy by} & \textbf{Std of} & \textbf{Entropy by} & \textbf{Number of} & \textbf{Std of} & \textbf{Std of} \\
& \textbf{brand} & \textbf{brands' SES} & \textbf{brands' price} & \textbf{brands} & \textbf{brands'} & \textbf{brands'} \\
& & & \textbf{level} & & \textbf{SES} & \textbf{price level} \\
\midrule
Amusement, Gambling, and & 0.278*** & 0.218*** & 0.082*** & 89 & 19314 & 1.035 \\
Recreation Industries & & & & & & \\
Clothing and Clothing & 0.142*** & 0.049*** & 0.223*** & 203 & 19654 & 0.768 \\
Accessories Stores & & & & & & \\
Food Services and Drinking & 0.141*** & 0.347*** & 0.430*** & 297 & 18584 & 0.663 \\
Places & & & & & & \\
Miscellaneous Store & 0.130*** & 0.100*** & 0.013 & 49 & 23878 & 0.598 \\
Retailers & & & & & & \\
Sporting Goods, Hobby, & 0.107*** & 0.143*** & 0.079*** & 53 & 25145 & 0.592 \\
Musical Instrument, and & & & & & & \\
Book Stores & & & & & & \\
Motion Picture and Video & 0.046* & 0.015 & -- & 9 & 21571 & -- \\
Industries & & & & & & \\
Health and Personal Care & 0.034*** & 0.084*** & 0.089*** & 47 & 14221 & 0.494 \\
Stores & & & & & & \\
Food and Beverage Stores & -0.064*** & 0.035*** & -0.001 & 95 & 30776 & 0.640 \\
Gasoline Stations & -0.120*** & -0.090*** & -0.071*** & 35 & 20154 & 0.512 \\
General Merchandise Stores & -0.154*** & 0.055*** & 0.112*** & 42 & 21078 & 0.935 \\
\bottomrule
\end{tabular}
\begin{flushleft}
\footnotesize
Note: * $p < 0.05$, ** $p < 0.01$, *** $p < 0.001$ (two-tailed tests). Industries are ordered in descending order by entropy by brand. Yelp price levels are not available for brands in \emph{Motion Picture and Video Industries}. Excluded \emph{Personal and Laundry Services} and \emph{Rental and Leasing Services} due to limited number of brands.
\end{flushleft}
\end{table}

\begin{table}[h]
\centering
\caption{The associations between CBGs' proportion of brand visitors with bachelor's or higher degree and diversity in consumption by CBGs in and outside New York City (NYC).}
\label{tab:tab_e2}
\footnotesize
\setlength{\tabcolsep}{9pt}
\begin{tabular}{llrrr}
\toprule
\multicolumn{2}{l}{\textbf{Region}} & \textbf{Entropy by} & \textbf{Standard} & \textbf{Entropy by} \\
& & \textbf{brand} & \textbf{deviation of} & \textbf{brands' price} \\
& & & \textbf{brands' SES} & \textbf{level} \\
\midrule
\multirow{2}{*}{NYC CBGs} & visiting all stores & -0.155*** & 0.171*** & 0.262*** \\
& visiting only NYC stores & 0.287*** & 0.089*** & 0.103*** \\
\multirow{2}{*}{Non-NYC CBGs} & visiting all stores & 0.309*** & 0.494*** & 0.498*** \\
& visiting only non-NYC stores & 0.211*** & 0.490*** & 0.470*** \\
\bottomrule
\end{tabular}
\begin{flushleft}
\footnotesize
Note: * $p < 0.05$, ** $p < 0.01$, *** $p < 0.001$ (two-tailed tests).
\end{flushleft}
\end{table}

\begin{table}[ht]
\centering
\caption{Results from regression analyses that predict CBGs' diversity in consumption with median household income, proportion of bachelor's degree or higher, the interaction between income and education, income variability, estimated mobility, estimated local availability, and demographic variables.}
\label{tab:tab_e3}
\footnotesize
\setlength{\tabcolsep}{6pt}
\begin{tabular}{lrrr}
\toprule
& \textbf{Entropy by} & \textbf{Standard} & \textbf{Entropy by} \\
& \textbf{brand} & \textbf{deviation of} & \textbf{brands' price} \\
& & \textbf{brands' SES} & \textbf{level} \\
\midrule
Income & 0.698*** & 0.403*** & 0.263*** \\
& (0.019) & (0.022) & (0.019) \\
Proportion of & -0.145*** & 0.153*** & 0.218*** \\
bachelor's degree & (0.016) & (0.018) & (0.016) \\
or higher & & & \\
Income-education & -0.208*** & -0.066*** & -0.061*** \\
interaction & (0.010) & (0.012) & (0.010) \\
Income variability & -0.145*** & 0.039* & 0.017 \\
& (0.014) & (0.015) & (0.014) \\
Mobility & -0.199*** & -0.164*** & -0.082*** \\
& (0.013) & (0.015) & (0.013) \\
Local availability & 0.167*** & 0.228*** & 0.089*** \\
& (0.011) & (0.013) & (0.011) \\
In NYC & -0.040 & -0.466*** & 0.522*** \\
& (0.030) & (0.035) & (0.030) \\
Median age & 0.019 & 0.022 & 0.051*** \\
& (0.012) & (0.013) & (0.012) \\
Proportion of male & -0.033** & -0.024* & 0.007 \\
& (0.011) & (0.012) & (0.011) \\
Proportion of white & -0.193*** & -0.337*** & -0.125*** \\
& (0.015) & (0.018) & (0.015) \\
Intercept & 0.165*** & 0.216*** & -0.140*** \\
& (0.016) & (0.018) & (0.016) \\
\midrule
Observations & 6088 & 3672 & 5739 \\
R² & 0.295 & 0.406 & 0.319 \\
Adjusted R² & 0.294 & 0.405 & 0.318 \\
Residual Std. Error & 0.840 & 0.742 & 0.818 \\
F Statistic & 254.038*** & 250.596*** & 268.551*** \\
\bottomrule
\end{tabular}
\begin{flushleft}
\footnotesize
Note: * $p < 0.05$, ** $p < 0.01$, *** $p < 0.001$ (two-tailed tests).
\end{flushleft}
\end{table}

\begin{table}[ht]
\centering
\caption{Results from regression analyses that predict CBGs' diversity (the entropy by brand) by industry.}
\label{tab:tab_e4}
\tiny
\setlength{\tabcolsep}{3pt}
\begin{tabular}{lrrrrrrrrr}
\toprule
& \textbf{Amusement,} & \textbf{Clothing and} & \textbf{Food Services} & \textbf{Food and} & \textbf{Gasoline} & \textbf{General} & \textbf{Health and} & \textbf{Miscellaneous} & \textbf{Sporting} \\
& \textbf{Gambling,} & \textbf{Clothing} & \textbf{and Drinking} & \textbf{Beverage} & \textbf{Stations} & \textbf{Merchandise} & \textbf{Personal Care} & \textbf{Store} & \textbf{Goods,} \\
& \textbf{and Recreation} & \textbf{Accessories} & \textbf{Places} & \textbf{Stores} & & \textbf{Stores} & \textbf{Stores} & \textbf{Retailers} & \textbf{Hobby} \\
& \textbf{Industries} & \textbf{Stores} & & & & & & & \\
\midrule
Income & 0.533*** & 0.521*** & 0.543*** & 0.374*** & 0.494*** & 0.377*** & 0.285*** & 0.519*** & 0.454*** \\
& (0.051) & (0.070) & (0.026) & (0.035) & (0.026) & (0.040) & (0.043) & (0.057) & (0.085) \\
Proportion of & 0.093* & -0.007 & -0.079*** & -0.152*** & -0.235*** & -0.092** & -0.075* & -0.029 & 0.109 \\
bachelor's degree & (0.042) & (0.060) & (0.022) & (0.031) & (0.024) & (0.033) & (0.035) & (0.051) & (0.067) \\
or higher & & & & & & & & & \\
Income-education & -0.134*** & -0.134*** & -0.160*** & -0.067*** & -0.148*** & -0.203*** & -0.033 & -0.180*** & -0.183*** \\
interaction & (0.027) & (0.035) & (0.014) & (0.019) & (0.017) & (0.023) & (0.022) & (0.030) & (0.042) \\
Income variability & -0.119*** & -0.117** & -0.107*** & -0.095*** & -0.084*** & -0.051 & -0.088** & -0.137*** & -0.231*** \\
& (0.034) & (0.042) & (0.018) & (0.025) & (0.021) & (0.033) & (0.029) & (0.038) & (0.060) \\
Mobility & 0.013 & -0.074 & -0.104*** & 0.046* & 0.123*** & -0.038 & 0.171*** & 0.106 & 0.021 \\
& (0.051) & (0.042) & (0.020) & (0.022) & (0.023) & (0.026) & (0.034) & (0.056) & (0.067) \\
Local availability & 0.145*** & 0.206*** & 0.192*** & 0.029 & 0.110*** & 0.065** & 0.155*** & 0.131*** & 0.106* \\
& (0.030) & (0.040) & (0.015) & (0.020) & (0.015) & (0.020) & (0.024) & (0.035) & (0.048) \\
In NYC & 0.314*** & 0.558*** & 0.190*** & -0.130* & -0.417*** & -0.462*** & 0.345*** & -0.450*** & -0.042 \\
& (0.080) & (0.111) & (0.042) & (0.058) & (0.049) & (0.069) & (0.065) & (0.099) & (0.134) \\
Median age & 0.006 & 0.060 & -0.011 & -0.004 & -0.003 & 0.025 & -0.047 & 0.000 & 0.150** \\
& (0.030) & (0.041) & (0.016) & (0.021) & (0.016) & (0.021) & (0.025) & (0.036) & (0.047) \\
Proportion of male & -0.017 & 0.019 & -0.025 & -0.029 & -0.035* & -0.041* & -0.048 & -0.006 & 0.073 \\
& (0.028) & (0.039) & (0.015) & (0.020) & (0.016) & (0.020) & (0.025) & (0.034) & (0.044) \\
Proportion of white & -0.104* & -0.147* & -0.169*** & -0.149*** & -0.141*** & -0.099*** & -0.033 & -0.049 & -0.168* \\
& (0.045) & (0.063) & (0.021) & (0.028) & (0.022) & (0.030) & (0.034) & (0.050) & (0.072) \\
Intercept & 0.215*** & 0.132* & 0.159*** & 0.184*** & 0.427*** & 0.390*** & 0.004 & 0.569*** & 0.449*** \\
& (0.043) & (0.064) & (0.022) & (0.029) & (0.021) & (0.031) & (0.035) & (0.045) & (0.066) \\
\midrule
Observations & 814 & 523 & 3192 & 2071 & 2559 & 1217 & 1578 & 608 & 362 \\
R² & 0.286 & 0.237 & 0.223 & 0.064 & 0.187 & 0.175 & 0.098 & 0.232 & 0.181 \\
Adjusted R² & 0.277 & 0.222 & 0.221 & 0.059 & 0.183 & 0.168 & 0.093 & 0.219 & 0.158 \\
Residual Std. Error & 0.816 & 0.914 & 0.840 & 0.891 & 0.752 & 0.677 & 0.942 & 0.834 & 0.894 \\
F Statistic & 32.213*** & 15.917*** & 91.408*** & 14.074*** & 58.456*** & 25.497*** & 17.111*** & 18.040*** & 7.774*** \\
\bottomrule
\end{tabular}
\begin{flushleft}
\footnotesize
Note: *** $p < 0.001$, ** $p < 0.01$, * $p < 0.05$ (two-tailed tests). No regression model for the industry \emph{Motion Picture and Video Industries} due to limited number of observations with price and SES data.
\end{flushleft}
\end{table}

\begin{table}[ht]
\centering
\caption{Results from regression analyses that predict CBGs' diversity (standard deviation of brands' SES) by industry.}
\label{tab:tab_e5}
\tiny
\setlength{\tabcolsep}{3pt}
\begin{tabular}{lrrrrrrrrr}
\toprule
& \textbf{Amusement,} & \textbf{Clothing and} & \textbf{Food Services} & \textbf{Food and} & \textbf{Gasoline} & \textbf{General} & \textbf{Health and} & \textbf{Miscellaneous} & \textbf{Sporting} \\
& \textbf{Gambling,} & \textbf{Clothing} & \textbf{and Drinking} & \textbf{Beverage} & \textbf{Stations} & \textbf{Merchandise} & \textbf{Personal Care} & \textbf{Store} & \textbf{Goods,} \\
& \textbf{and Recreation} & \textbf{Accessories} & \textbf{Places} & \textbf{Stores} & & \textbf{Stores} & \textbf{Stores} & \textbf{Retailers} & \textbf{Hobby} \\
& \textbf{Industries} & \textbf{Stores} & & & & & & & \\
\midrule
Income & 0.365** & 0.273** & 0.354*** & 0.276*** & 0.016 & 0.229** & 0.311*** & 0.229* & 0.451*** \\
& (0.111) & (0.094) & (0.035) & (0.069) & (0.055) & (0.073) & (0.086) & (0.103) & (0.126) \\
Proportion of & 0.214* & -0.015 & 0.077** & -0.022 & 0.060 & 0.011 & -0.089 & -0.027 & 0.002 \\
bachelor's degree & (0.083) & (0.079) & (0.028) & (0.063) & (0.049) & (0.060) & (0.070) & (0.101) & (0.114) \\
or higher & & & & & & & & & \\
Income-education & -0.196*** & -0.082 & -0.108*** & 0.000 & -0.033 & -0.041 & -0.032 & 0.002 & -0.158 \\
interaction & (0.057) & (0.047) & (0.018) & (0.042) & (0.037) & (0.043) & (0.037) & (0.062) & (0.098) \\
Income variability & 0.028 & 0.016 & 0.065** & -0.032 & -0.060 & 0.110 & -0.092 & 0.152* & -0.101 \\
& (0.060) & (0.057) & (0.024) & (0.048) & (0.044) & (0.061) & (0.055) & (0.068) & (0.082) \\
Mobility & 0.252* & -0.061 & -0.034 & 0.092 & 0.035 & 0.030 & 0.208** & 0.053 & 0.405** \\
& (0.126) & (0.069) & (0.027) & (0.065) & (0.056) & (0.059) & (0.076) & (0.126) & (0.122) \\
Local availability & 0.140* & 0.125* & 0.278*** & 0.264*** & 0.368*** & 0.091* & 0.272*** & 0.176** & 0.354*** \\
& (0.056) & (0.058) & (0.020) & (0.040) & (0.031) & (0.039) & (0.049) & (0.064) & (0.075) \\
In NYC & 0.265 & 0.224 & -0.325*** & -0.223 & -0.294** & 0.024 & 0.075 & -0.324 & -0.130 \\
& (0.144) & (0.150) & (0.056) & (0.147) & (0.105) & (0.160) & (0.132) & (0.252) & (0.238) \\
Median age & -0.024 & 0.006 & 0.058** & 0.038 & 0.026 & 0.066 & -0.038 & 0.016 & 0.143* \\
& (0.056) & (0.054) & (0.021) & (0.038) & (0.036) & (0.039) & (0.046) & (0.062) & (0.069) \\
Proportion of male & 0.054 & -0.119* & -0.021 & 0.047 & 0.003 & -0.009 & -0.031 & 0.083 & 0.106 \\
& (0.052) & (0.052) & (0.019) & (0.046) & (0.034) & (0.043) & (0.045) & (0.067) & (0.069) \\
Proportion of white & -0.116 & -0.270** & -0.220*** & -0.456*** & -0.124** & -0.253*** & 0.007 & -0.126 & -0.181 \\
& (0.092) & (0.082) & (0.028) & (0.069) & (0.046) & (0.068) & (0.074) & (0.107) & (0.111) \\
Intercept & 0.182* & 0.188* & 0.311*** & -0.018 & 0.346*** & 0.091 & 0.170* & 0.369*** & 0.423*** \\
& (0.085) & (0.086) & (0.028) & (0.060) & (0.042) & (0.058) & (0.073) & (0.085) & (0.101) \\
\midrule
Observations & 150 & 246 & 1651 & 295 & 684 & 281 & 385 & 147 & 98 \\
R² & 0.334 & 0.167 & 0.299 & 0.365 & 0.187 & 0.210 & 0.147 & 0.237 & 0.381 \\
Adjusted R² & 0.286 & 0.132 & 0.295 & 0.343 & 0.174 & 0.181 & 0.124 & 0.181 & 0.310 \\
Residual Std. Error & 0.634 & 0.839 & 0.794 & 0.652 & 0.804 & 0.642 & 0.892 & 0.730 & 0.699 \\
F Statistic & 6.970*** & 4.715*** & 69.942*** & 16.346*** & 15.434*** & 7.194*** & 6.421*** & 4.223*** & 5.356*** \\
\bottomrule
\end{tabular}
\begin{flushleft}
\footnotesize
Note: *** $p < 0.001$, ** $p < 0.01$, * $p < 0.05$ (two-tailed tests). No regression model for the industry \emph{Motion Picture and Video Industries} due to limited number of observations with price and SES data.
\end{flushleft}
\end{table}

\begin{table}[ht]
\centering
\caption{Results from regression analyses that predict CBGs' diversity (entropy by brands' price level) by industry.}
\label{tab:tab_e6}
\tiny
\setlength{\tabcolsep}{3pt}
\begin{tabular}{lrrrrrrrrr}
\toprule
& \textbf{Amusement,} & \textbf{Clothing and} & \textbf{Food Services} & \textbf{Food and} & \textbf{Gasoline} & \textbf{General} & \textbf{Health and} & \textbf{Miscellaneous} & \textbf{Sporting} \\
& \textbf{Gambling,} & \textbf{Clothing} & \textbf{and Drinking} & \textbf{Beverage} & \textbf{Stations} & \textbf{Merchandise} & \textbf{Personal Care} & \textbf{Store} & \textbf{Goods,} \\
& \textbf{and Recreation} & \textbf{Accessories} & \textbf{Places} & \textbf{Stores} & & \textbf{Stores} & \textbf{Stores} & \textbf{Retailers} & \textbf{Hobby} \\
& \textbf{Industries} & \textbf{Stores} & & & & & & & \\
\midrule
Income & 0.332* & 0.179* & 0.267*** & 0.226*** & 0.366*** & 0.417*** & -0.123** & 0.120 & 0.075 \\
& (0.157) & (0.071) & (0.026) & (0.035) & (0.035) & (0.042) & (0.046) & (0.066) & (0.099) \\
Proportion of & 0.050 & 0.196** & 0.283*** & -0.035 & -0.206*** & 0.112** & 0.202*** & -0.074 & 0.224** \\
bachelor's degree & (0.127) & (0.060) & (0.022) & (0.031) & (0.031) & (0.036) & (0.038) & (0.058) & (0.079) \\
or higher & & & & & & & & & \\
Income-education & -0.034 & -0.073* & -0.108*** & -0.027 & -0.053* & -0.160*** & 0.088*** & -0.024 & -0.043 \\
interaction & (0.080) & (0.036) & (0.014) & (0.019) & (0.022) & (0.025) & (0.023) & (0.035) & (0.050) \\
Income variability & -0.164 & 0.006 & 0.007 & -0.061** & -0.026 & -0.013 & -0.020 & -0.100** & -0.127 \\
& (0.124) & (0.042) & (0.018) & (0.025) & (0.027) & (0.035) & (0.031) & (0.044) & (0.070) \\
Mobility & -0.020 & -0.005 & 0.019 & -0.003 & 0.006 & -0.005 & 0.130*** & -0.056 & 0.011 \\
& (0.142) & (0.043) & (0.019) & (0.022) & (0.030) & (0.028) & (0.037) & (0.064) & (0.075) \\
Local availability & 0.165 & 0.166*** & 0.130*** & 0.079*** & 0.176*** & 0.104*** & 0.155*** & 0.162*** & 0.172** \\
& (0.100) & (0.041) & (0.015) & (0.020) & (0.019) & (0.021) & (0.027) & (0.040) & (0.056) \\
In NYC & -0.070 & 0.576*** & 0.227*** & -0.441*** & 0.286*** & 0.195** & 0.343*** & -0.277* & 0.158 \\
& (0.281) & (0.110) & (0.041) & (0.057) & (0.062) & (0.074) & (0.070) & (0.112) & (0.154) \\
Median age & -0.077 & 0.080 & 0.034* & -0.041 & 0.009 & 0.072** & -0.005 & 0.002 & 0.119* \\
& (0.091) & (0.042) & (0.016) & (0.021) & (0.022) & (0.022) & (0.027) & (0.041) & (0.057) \\
Proportion of male & -0.057 & 0.031 & 0.021 & -0.007 & 0.006 & 0.010 & 0.083** & -0.037 & 0.011 \\
& (0.092) & (0.040) & (0.014) & (0.020) & (0.021) & (0.021) & (0.026) & (0.039) & (0.052) \\
Proportion of white & 0.079 & -0.033 & -0.026 & 0.066* & -0.173*** & -0.107*** & 0.042 & -0.041 & -0.255** \\
& (0.143) & (0.063) & (0.020) & (0.028) & (0.028) & (0.032) & (0.036) & (0.058) & (0.084) \\
Intercept & 0.244 & -0.052 & 0.021 & 0.267*** & 0.021 & 0.101** & -0.166*** & 0.272*** & 0.059 \\
& (0.125) & (0.064) & (0.021) & (0.029) & (0.027) & (0.033) & (0.037) & (0.052) & (0.075) \\
\midrule
Observations & 178 & 512 & 3191 & 1980 & 2098 & 1195 & 1442 & 539 & 330 \\
R² & 0.091 & 0.238 & 0.272 & 0.120 & 0.154 & 0.266 & 0.127 & 0.079 & 0.124 \\
Adjusted R² & 0.036 & 0.223 & 0.269 & 0.115 & 0.150 & 0.260 & 0.121 & 0.062 & 0.097 \\
Residual Std. Error & 1.116 & 0.910 & 0.825 & 0.875 & 0.871 & 0.715 & 0.970 & 0.912 & 0.981 \\
F Statistic & 1.665 & 15.659*** & 118.570*** & 26.787*** & 37.862*** & 42.891*** & 20.850*** & 4.526*** & 4.535*** \\
\bottomrule
\end{tabular}
\begin{flushleft}
\footnotesize
Note: *** $p < 0.001$, ** $p < 0.01$, * $p < 0.05$ (two-tailed tests). No regression model for the industry \emph{Motion Picture and Video Industries} due to limited number of observations with price and SES data.
\end{flushleft}
\end{table}

\clearpage

\section{Replicating the analyses for Texas}
\label{app:appendix_f}

\setcounter{figure}{0}
\setcounter{table}{0}
\renewcommand{\thefigure}{\Alph{section}\arabic{figure}}
\renewcommand{\thetable}{\Alph{section}\arabic{table}}

\subsection{Data}

\begin{figure}[ht]
\centering
\includegraphics[width=0.7\textwidth]{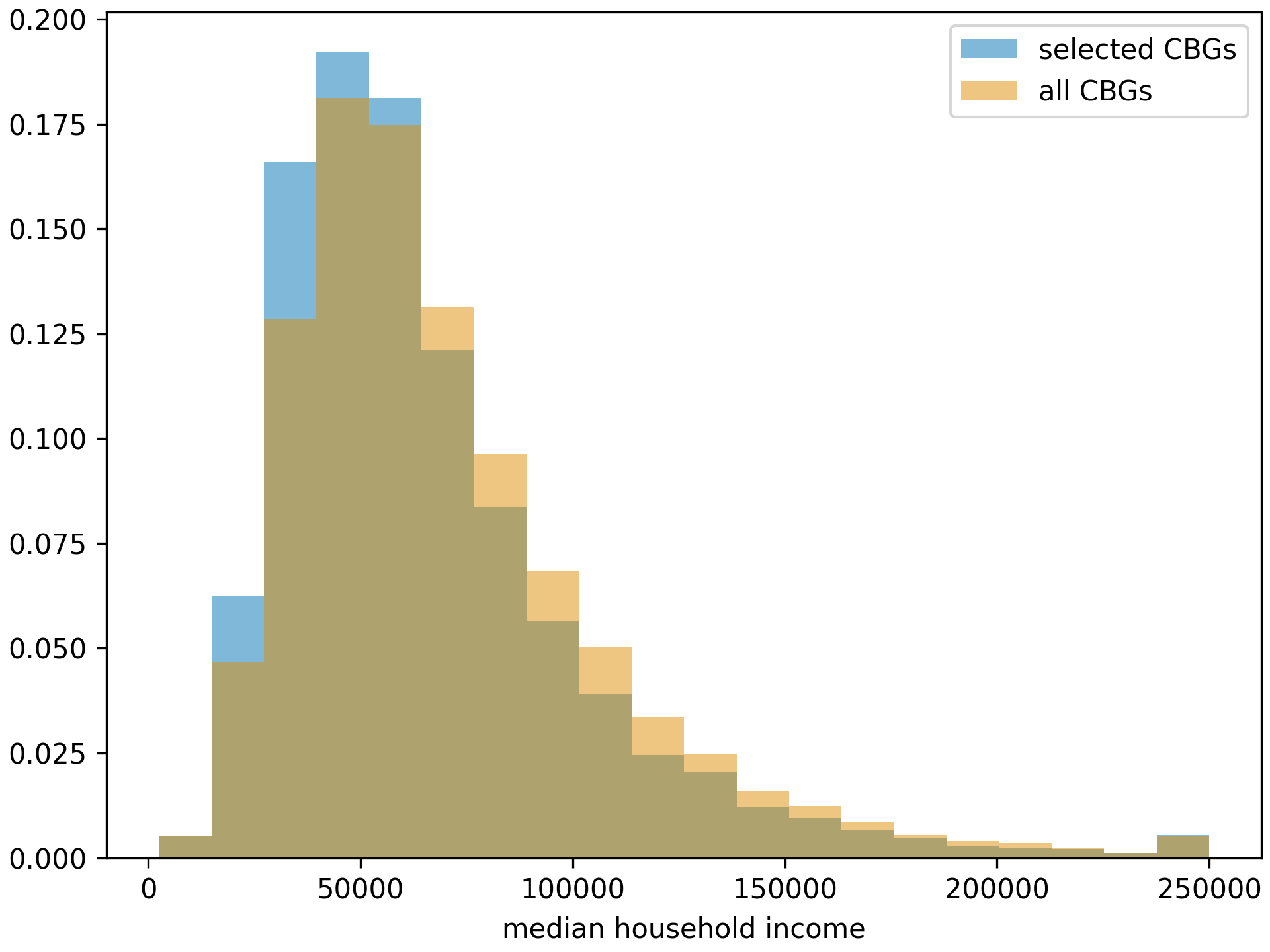}
\caption{The income distribution for selected and all CBGs in the Texas data.}
\label{fig:fig_f1}
\end{figure}

After filtering the 28 four-digit NAICS codes, there are 323,098 observations in total for the Texas data in October 2019. Table \ref{tab:tab_f1} shows the four-digit category names, the NAICS codes, and the number of POIs for each category. Among the 323,098 observations, 83,299 (about 26 percent) have brands associated with them. After dropping observations with limited information, we have 56,266 POIs that belong to 1,484 brands. Among the 28 categories, two of them (\emph{Gambling Industries} and \emph{Museums, Historical Sites, and Similar Institutions}), do not have any brand, so we end up with 26 categories. For 55,268 of the 56,266 POIs, each observation has a table of the visitors' home CBGs and the number of visitors from each CBG. The visitors to the 55,268 POIs are from 66,523 CBGs, where 64,753 median household income estimates are available. Aggregating the data by brands, we get a bipartite network of 64,753 CBGs and 1,459 brands, where the weights are the number of visits from each CBG to each brand. We drop the brands that have fewer than 100 incoming visitors and CBGs that have fewer than 100 outgoing visitors, resulting in a network of 15,729 CBGs and 1,273 brands. As shown in Figure~\ref{fig:fig_f1}, compared with all 64,753 CBGs, the selected 15,729 CBGs slightly oversample lower income CBGs. For the 1,273 brands, we were able to match 511 brands with the Yelp Open Dataset and manually found the Yelp price level for 762 brands, leaving 265 brands missing.

\newpage

\begin{table}[ht]
\centering
\caption{Four-digit NAICS industries and number of POIs included in the Texas data.}
\label{tab:tab_f1}
\footnotesize
\begin{tabular}{lrr}
\toprule
\textbf{Four-digit industry name} & \textbf{NAICS codes} & \textbf{Number of POIs} \\
\midrule
Restaurants and Other Eating Places & 7225 & 97658 \\
Personal Care Services & 8121 & 55192 \\
Gasoline Stations & 4471 & 19588 \\
Other Amusement and Recreation Industries & 7139 & 17889 \\
Grocery Stores & 4451 & 17377 \\
Health and Personal Care Stores & 4461 & 15364 \\
Clothing Stores & 4481 & 12814 \\
Museums, Historical Sites, and Similar Institutions & 7121 & 11073 \\
Other Miscellaneous Store Retailers & 4539 & 9237 \\
Drinking Places (Alcoholic Beverages) & 7224 & 8215 \\
Sporting Goods, Hobby, and Musical Instrument Stores & 4511 & 7884 \\
Consumer Goods Rental & 5322 & 7586 \\
Specialty Food Stores & 4452 & 5900 \\
Jewelry, Luggage, and Leather Goods Stores & 4483 & 5872 \\
General Merchandise Stores, including Warehouse Clubs and Supercenters & 4523 & 5477 \\
Florists & 4531 & 4098 \\
Used Merchandise Stores & 4533 & 4075 \\
Beer, Wine, and Liquor Stores & 4453 & 4036 \\
Office Supplies, Stationery, and Gift Stores & 4532 & 3362 \\
Shoe Stores & 4482 & 2587 \\
Amusement Parks and Arcades & 7131 & 1614 \\
Special Food Services & 7223 & 1508 \\
Department Stores & 4522 & 1495 \\
Book Stores and News Dealers & 4512 & 1387 \\
Gambling Industries & 7132 & 693 \\
Motion Picture and Video Industries & 5121 & 692 \\
Spectator Sports & 7112 & 232 \\
Performing Arts Companies & 7111 & 193 \\
\bottomrule
\end{tabular}
\end{table}

\subsection{Descriptive results}

\begin{figure}[ht]
\centering
\includegraphics[width=\textwidth]{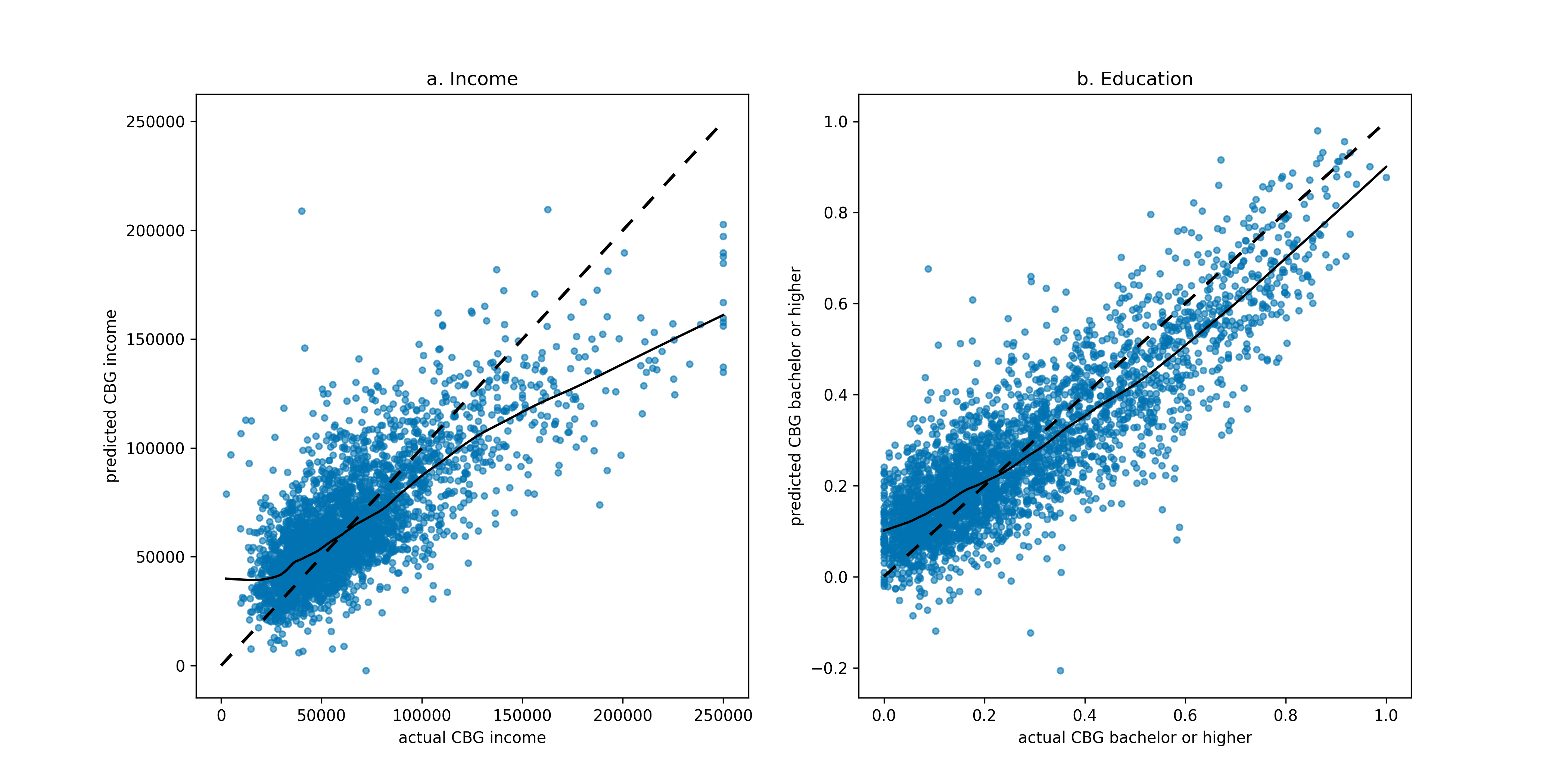}
\caption{Correlation between predicted and actual income or education in the test set (Texas). The solid lines show the LOESS fit and the dashed lines -- the ideal 1:1 relation.}
\label{fig:fig_f2}
\end{figure}

Using simple LASSO regression models, we can predict the median household income or the proportion of people who have bachelor's or higher degree of the CBGs with the proportion of outgoing visitors of the CBGs to the brands. The out-of-sample correlations are 0.753 for income and 0.859 for education, both statistically significant at the 0.001 level. The models explain 0.640 variance for income and 0.767 variance for education. Figure~\ref{fig:fig_f2} shows the correlation between predicted and actual income (panel a) or education (panel b) in the test set.

\begin{figure}[ht]
\centering
\includegraphics[width=\textwidth]{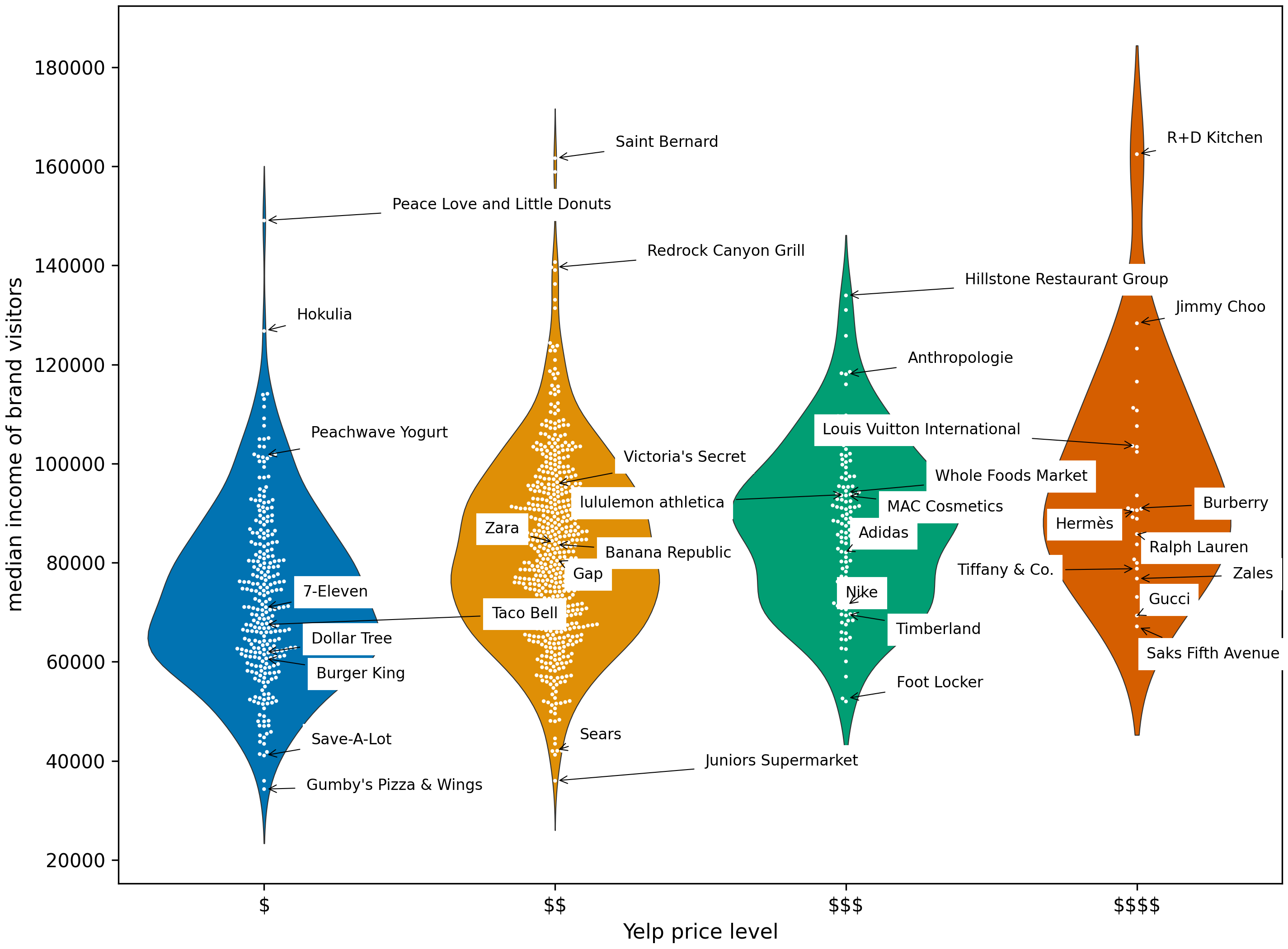}
\caption{Relation between brands’ Yelp price level and median income of brand visitors (Texas). One extreme outlier The Pizza Press (median income 9,222; yelp price level \$) is excluded for better visualisation.} 
\label{fig:fig_f3}
\end{figure}

\begin{figure}[ht]
\centering
\includegraphics[width=\textwidth]{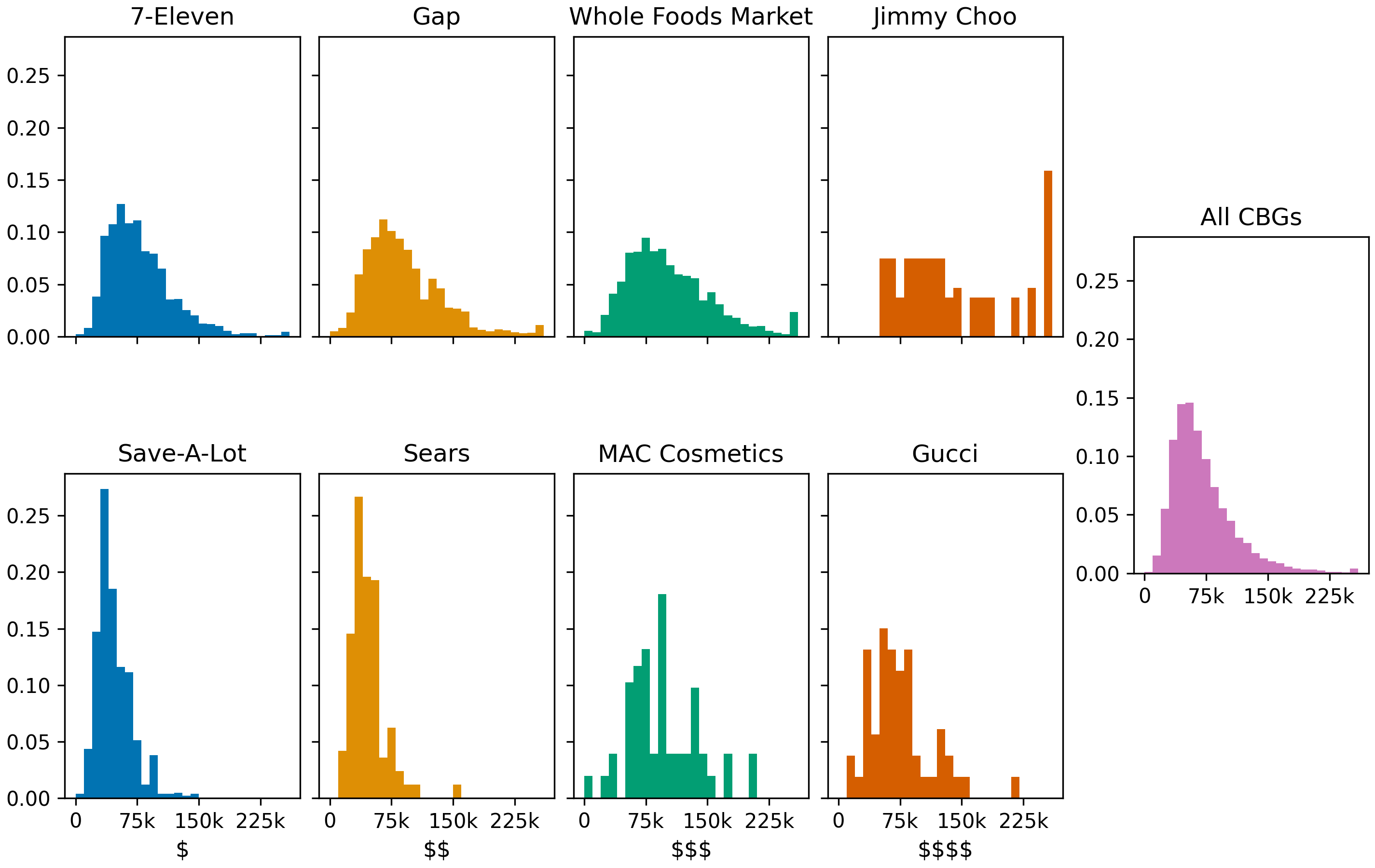}
\caption{Distribution of brand visitors’ income for some typical brands and all CBGs in our sample (Texas).}
\label{fig:fig_f4}
\end{figure}

\begin{figure}[ht]
\centering
\includegraphics[width=\textwidth]{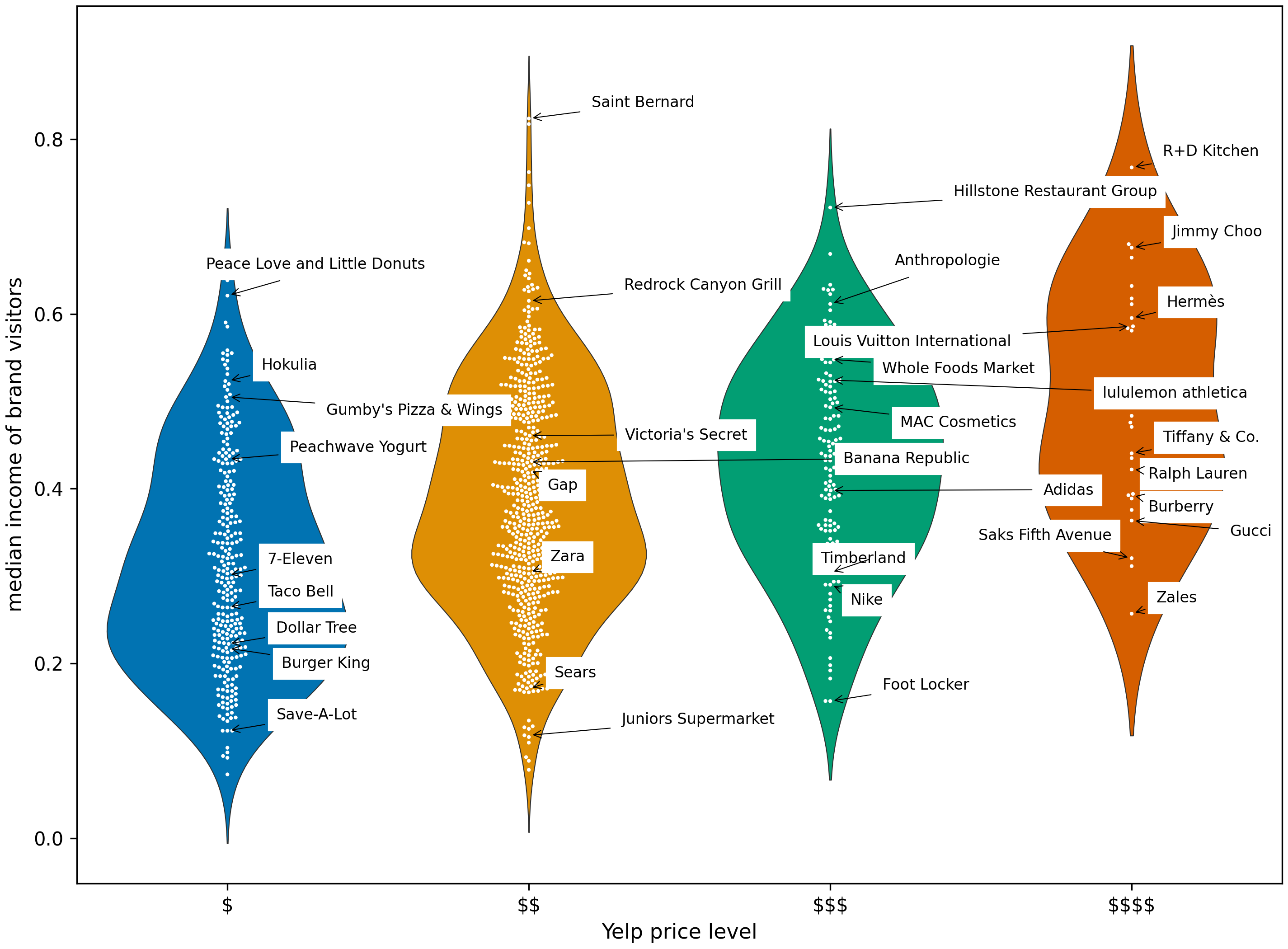}
\caption{Relation between brands’ Yelp price level and median proportion of brand visitors with bachelor’s or higher degree (Texas).}
\label{fig:fig_f5}
\end{figure}

\begin{figure}[ht]
\centering
\includegraphics[width=\textwidth]{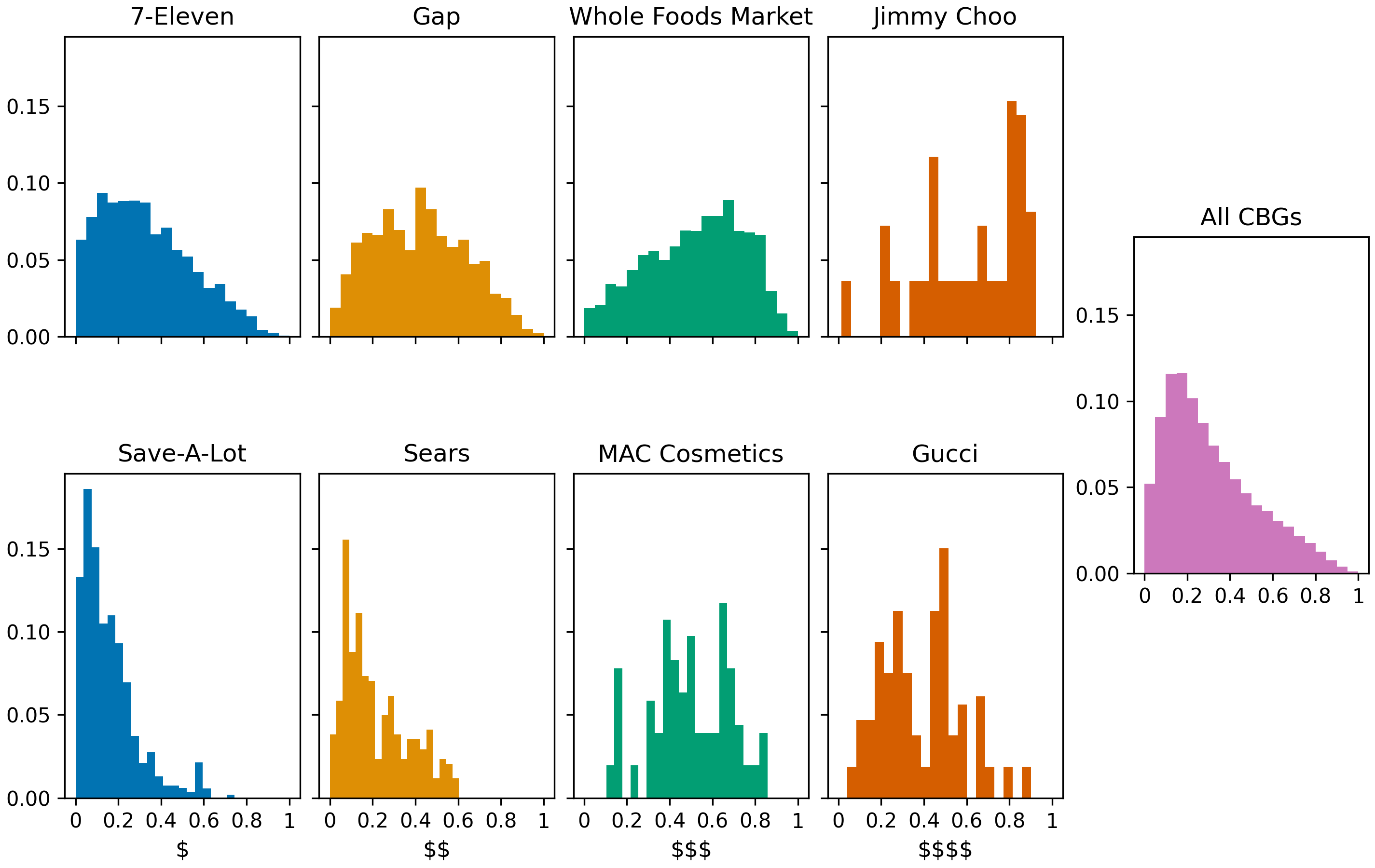}
\caption{Distribution of proportion of brand visitors with bachelor’s or higher degree for typical brands and all CBGs in our sample (Texas).}
\label{fig:fig_f6}
\end{figure}

Figure~\ref{fig:fig_f3} shows the distribution of brands’ SES (measured by the median income of brand visitors) for different Yelp price levels, with some typical brands labelled. Figure~\ref{fig:fig_f4} shows the distribution of brand visitors’ income for some typical brands identified from Figure~\ref{fig:fig_f3}, and for reference, the distribution of median household income for the CBGs in our sample. We try to use the same brands as in New York State, but Lidl and Valentino do not have data available in Texas, so they are replaced with similar brands 7-Eleven and Jimmy Choo. The patterns are similar with what we find in New York State. Similar patterns exist for education, as shown in Figure~\ref{fig:fig_f5} and Figure~\ref{fig:fig_f6}.

\clearpage

\subsection{Diversity in consumption}

\begin{figure}[ht]
\centering
\includegraphics[width=\textwidth]{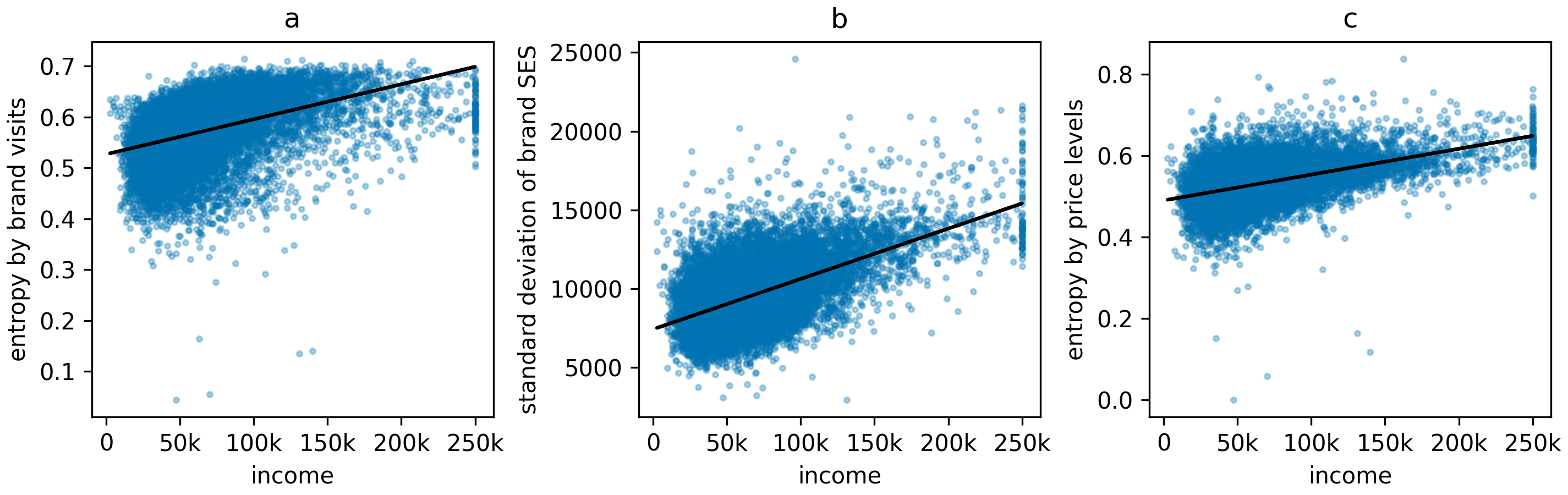}
\caption{Correlation between CBGs’ median household income and three measures of diversity (Texas).}
\label{fig:fig_f7}
\end{figure}

\begin{figure}[ht]
\centering
\includegraphics[width=\textwidth]{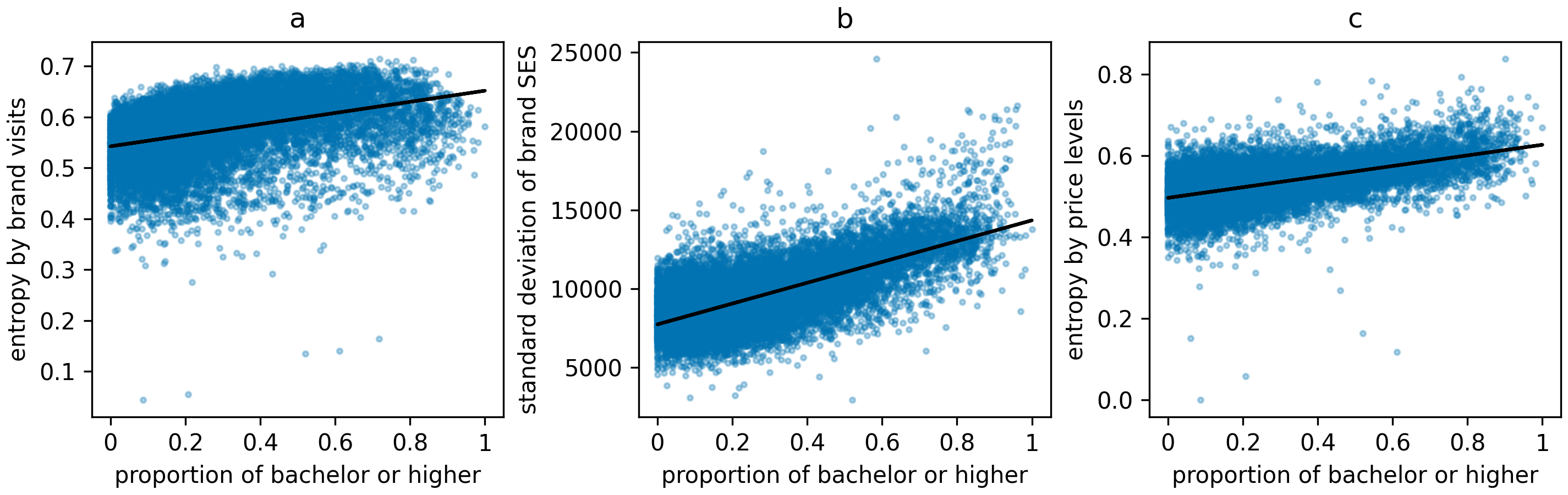}
\caption{Correlation between CBGs’ proportion of residents with bachelor’s or higher degree and three measures of diversity (Texas).}
\label{fig:fig_f8}
\end{figure}

The consumption diversity hypothesis is confirmed again with Texas data. As shown in Table \ref{tab:tab_f2}, all measures of diversity have significant and positive correlation with CBGs' median household income. Figure~\ref{fig:fig_f7} shows the correlations between CBGs' median household income and the three measures of diversity that we focus on reporting. The correlation coefficients are 0.433, 0.560, and 0.473 respectively for the entropy by brand, the standard deviation of visited brands' SES, and the entropy by brands' price level. Figure~\ref{fig:fig_f8} shows correlation between CBGs' proportion of residents with bachelor's or higher degree and the three measures of diversity. The correlation coefficients are 0.395, 0.663, and 0.557 respectively. Notably, the correlations in Texas are stronger than in New York State, especially for education, which we interpret to indicate higher consumption inequality in Texas.

\begin{table}[ht]
\centering
\caption{Correlations between median household income, proportion of residents who have a bachelor's or higher degree, and six different measures of diversity in consumption (Texas).}
\label{tab:tab_f2}
\footnotesize
\setlength{\tabcolsep}{4pt}
\begin{tabular}{lrrrrrrrr}
\toprule
& \textbf{Income} & \textbf{Proportion of} & \textbf{Number of} & \textbf{Brand} & \textbf{SES range} & \textbf{SES std} & \textbf{Number of} & \textbf{Price} \\
& & \textbf{bachelor or} & \textbf{brands} & \textbf{entropy} & & & \textbf{price level} & \textbf{entropy} \\
& & \textbf{higher} & & & & & & \\
\midrule
\textbf{Income} & 1 & 0.732*** & 0.409*** & 0.433*** & 0.432*** & 0.560*** & 0.322*** & 0.473*** \\
\textbf{Proportion of} & 0.732*** & 1 & 0.331*** & 0.395*** & 0.443*** & 0.663*** & 0.318*** & 0.557*** \\
\textbf{bachelor or higher} & & & & & & & & \\
\textbf{Number of brands} & 0.409*** & 0.331*** & 1 & 0.850*** & 0.603*** & 0.344*** & 0.530*** & 0.298*** \\
\textbf{Brand entropy} & 0.433*** & 0.395*** & 0.850*** & 1 & 0.649*** & 0.479*** & 0.547*** & 0.395*** \\
\textbf{SES range} & 0.432*** & 0.443*** & 0.603*** & 0.649*** & 1 & 0.727*** & 0.432*** & 0.372*** \\
\textbf{SES std} & 0.560*** & 0.663*** & 0.344*** & 0.479*** & 0.727*** & 1 & 0.349*** & 0.530*** \\
\textbf{Number of price level} & 0.322*** & 0.318*** & 0.530*** & 0.547*** & 0.432*** & 0.349*** & 1 & 0.594*** \\
\textbf{Price entropy} & 0.473*** & 0.557*** & 0.298*** & 0.395*** & 0.372*** & 0.530*** & 0.594*** & 1 \\
\bottomrule
\end{tabular}
\begin{flushleft}
\footnotesize
Note: *** $p < 0.001$ (two-tailed tests).
\end{flushleft}
\end{table}

Table \ref{tab:tab_f3} shows the association between CBGs’ median household income or CBGs’ proportion of residents with bachelor’s or higher degree and the diversity measures by industry. With only one exception of no correlation, there are significant associations between income or education and the diversity measures in all the industries, confirming the robustness of the consumption diversity hypothesis. The association is again stronger in industries that involve more cultural aspects than those that concern necessity goods. Compared with the results for New York State, where the associations between education and some measures of diversity are negative for a few industries involving necessity goods, the results in Texas provide an even stronger validation of our hypothesis. 

To explore urban-rural lifestyle differences, we disaggregate the analysis by Greater Houston and the rest of the state. We select Greater Houston because Houston is the largest and most populous city in Texas, but the distinction between Greater Houston and the rest of Texas is not as pronounced as New York City and the rest of New York State. We study the following cases: visits from CBGs in Greater Houston to stores in Texas, visits from CBGs outside of Greater Houston to stores in Texas, visits from CBGs in Greater Houston only to stores in Greater Houston, and visits from CBGs outside of Greater Houston to stores outside of Greater Houston. Table \ref{tab:tab_f4} shows the associations between CBGs’ income or education and the diversity measures for those cases. There are some differences in the associations between income or education and diversity between CBGs inside and outside of Greater Houston, but the differences are not as substantial as in New York City. It seems that New York City is a special case of weak association between socioeconomic status and consumption diversity.

\newpage

\begin{sidewaystable}[ht]
\centering
\caption{The associations between CBGs' median household income or CBGs' proportion residents with bachelor's or higher degree and diversity in consumption by industry.}
\label{tab:tab_f3}
\footnotesize
\setlength{\tabcolsep}{3pt}
\begin{tabular}{lrrrrrrrrrr}
\toprule
\textbf{Industry} & \multicolumn{3}{c}{\textbf{Associations with income}} & \multicolumn{3}{c}{\textbf{Associations with education}} & \multicolumn{3}{c}{\textbf{Industry characteristics}} \\
\cmidrule(lr){2-4} \cmidrule(lr){5-7} \cmidrule(lr){8-10}
& \textbf{Entropy by} & \textbf{Std of} & \textbf{Entropy by} & \textbf{Entropy by} & \textbf{Std of} & \textbf{Entropy by} & \textbf{Number of} & \textbf{Std of} & \textbf{Std of} \\
& \textbf{brand} & \textbf{brands' SES} & \textbf{brands' price} & \textbf{brand} & \textbf{brands' SES} & \textbf{brands' price} & \textbf{brands} & \textbf{brands'} & \textbf{brands'} \\
& & & \textbf{level} & & & \textbf{level} & & \textbf{SES} & \textbf{price level} \\
\midrule
Amusement, Gambling, and & 0.445*** & 0.280*** & 0.203*** & 0.432*** & 0.301*** & 0.179*** & 116 & 21331 & 0.667 \\
Recreation Industries & & & & & & & & & \\
Food Services and Drinking & 0.396*** & 0.514*** & 0.526*** & 0.362*** & 0.616*** & 0.624*** & 554 & 19026 & 0.619 \\
Places & & & & & & & & & \\
Miscellaneous Store & 0.344*** & 0.268*** & 0.156*** & 0.308*** & 0.249*** & 0.126*** & 61 & 18130 & 0.581 \\
Retailers & & & & & & & & & \\
Sporting Goods, Hobby, & 0.338*** & 0.237*** & 0.202*** & 0.292*** & 0.218*** & 0.260*** & 60 & 22591 & 0.572 \\
Musical Instrument, and & & & & & & & & & \\
Book Stores & & & & & & & & & \\
Clothing and Clothing & 0.319*** & 0.177*** & 0.269*** & 0.268*** & 0.182*** & 0.259*** & 210 & 17306 & 0.686 \\
Accessories Stores & & & & & & & & & \\
Health and Personal Care & 0.262*** & 0.249*** & 0.023** & 0.212*** & 0.222*** & 0.024** & 56 & 14842 & 0.436 \\
Stores & & & & & & & & & \\
General Merchandise & 0.203*** & 0.453*** & 0.397*** & 0.158*** & 0.524*** & 0.472*** & 41 & 16618 & 0.897 \\
Stores & & & & & & & & & \\
Motion Picture and Video & 0.189*** & 0.241*** & -- & 0.165*** & 0.210*** & -- & 14 & 24893 & -- \\
Industries & & & & & & & & & \\
Food and Beverage Stores & 0.177*** & 0.165*** & 0.115*** & 0.136*** & 0.214*** & 0.142*** & 98 & 18558 & 0.634 \\
Personal and Laundry & 0.139*** & 0.109** & 0.096* & 0.121*** & 0.081* & 0.088* & 16 & 19242 & 0.641 \\
Services & & & & & & & & & \\
Gasoline Stations & 0.136*** & 0.400*** & 0.031*** & 0.029*** & 0.400*** & -0.005 & 40 & 9205 & 0.512 \\
\bottomrule
\end{tabular}
\begin{flushleft}
\footnotesize
Note: * $p < 0.05$, ** $p < 0.01$, *** $p < 0.001$ (two-tailed tests). Industries are ordered in descending order by the association between income and entropy by brand. Yelp price levels are not available for brands in \emph{Motion Picture and Video Industries}. Excluded \emph{Performing Arts, Spectator Sports, and Related Industries} and \emph{Rental and Leasing Services} due to limited number of brands.
\end{flushleft}
\end{sidewaystable}

\newpage

\begin{sidewaystable}[ht]
\centering
\caption{The associations between CBGs' median household income and diversity in consumption by CBGs in and outside Greater Houston.}
\label{tab:tab_f4}
\footnotesize
\setlength{\tabcolsep}{6pt}
\begin{tabular}{llrrrrrr}
\toprule
\multicolumn{2}{l}{\textbf{Region}} & \multicolumn{3}{c}{\textbf{Associations with income}} & \multicolumn{3}{c}{\textbf{Associations with education}} \\
\cmidrule(lr){3-5} \cmidrule(lr){6-8}
& & \textbf{Entropy by} & \textbf{Std of} & \textbf{Entropy by} & \textbf{Entropy by} & \textbf{Std of} & \textbf{Entropy by} \\
& & \textbf{brand} & \textbf{brands' SES} & \textbf{brands' price} & \textbf{brand} & \textbf{brands' SES} & \textbf{brands' price} \\
& & & & \textbf{level} & & & \textbf{level} \\
\midrule
\multirow{2}{*}{Greater Houston CBGs} & visiting all stores & 0.392*** & 0.539*** & 0.539*** & 0.356*** & 0.707*** & 0.634*** \\
& visiting only Greater & 0.174*** & 0.421*** & 0.376*** & 0.272*** & 0.517*** & 0.455*** \\
& Houston stores & & & & & & \\
\multirow{2}{*}{Non-Greater Houston CBGs} & visiting all stores & 0.441*** & 0.577*** & 0.451*** & 0.406*** & 0.681*** & 0.539*** \\
& visiting only Non-Greater & 0.437*** & 0.575*** & 0.452*** & 0.401*** & 0.679*** & 0.539*** \\
& Houston stores & & & & & & \\
\bottomrule
\end{tabular}
\begin{flushleft}
\footnotesize
Note: *** $p < 0.001$ (two-tailed tests).
\end{flushleft}
\end{sidewaystable}

\subsection{Robustness to alternative explanations}

\begin{figure}[ht]
\centering
\includegraphics[width=\textwidth]{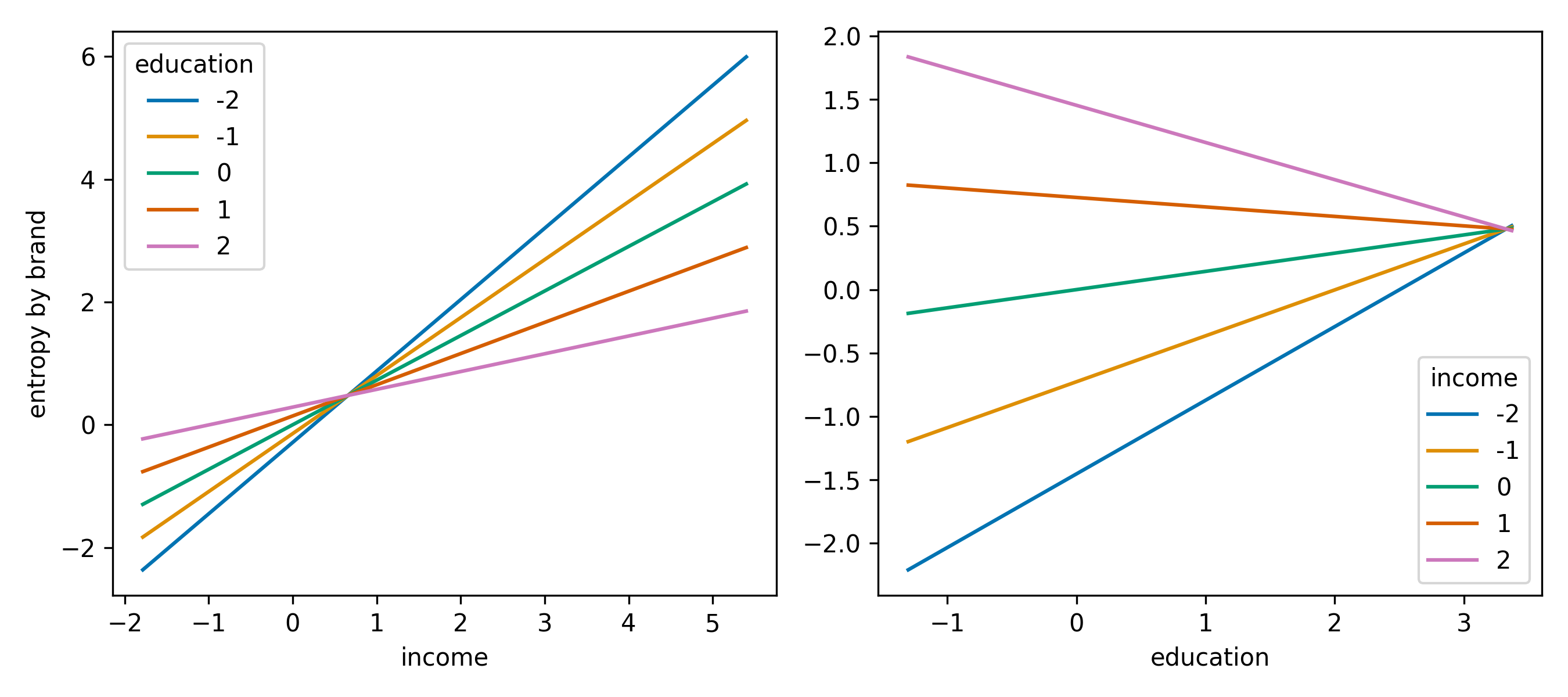}
\caption{The interaction effects between income and education on the entropy by brand (Texas).  The values are standardised. For example, -2 means two standard deviations lower than the mean value. The ranges of income and education on the x-axes are the actual ranges from the data.}
\label{fig:fig_f9}
\end{figure}

As in New York State, we use a standardised regression model to test the extent to which the associations between high SES and consumption diversity can be explained by simple geographic constraints. Table \ref{tab:tab_f5} shows the results from the regression models using different diversity measures. For all three measures of diversity, income and education are significantly associated with diversity, even controlling for income variability, mobility, local availability, age, gender, and race. Income is the predominant predictor for the entropy by brand, while education is the primary predictor for the standard deviation of brand’s SES and the entropy by brands’ price level. There is a significantly negative effect of the interaction between income and education for the entropy by brand. As shown in Figure \ref{fig:fig_f9}, the negative effect means that the association between income and entropy by brand visits is weaker for CBGs with higher education and the association between education and entropy by brand visits is negative for CBGs with higher income. There is a significantly positive but relatively weak effect of the interaction between income and education for the standard deviation of brands’ SES, meaning the association is stronger for the high income and high education group. The interaction effect is not significant for the entropy by brands’ price level. 

The findings are similar if we repeat the regression analyses separately by industry (see Tables \ref{tab:tab_f6}, \ref{tab:tab_f7}, and \ref{tab:tab_f8}). These results affirm the central hypothesis of this paper that high SES is associated with diverse consumption, but they also indicate nuances in the association that should be explored by future research with more refined data.

\begin{table}[ht]
\centering
\caption{Results from regression analyses that predict CBGs' diversity in consumption with median household income, proportion of bachelor's degree or higher, the interaction between income and education, income variability, estimated mobility, estimated local availability, and demographic variables (Texas).}
\label{tab:tab_f5}
\footnotesize
\setlength{\tabcolsep}{6pt}
\begin{tabular}{lrrr}
\toprule
& \textbf{Entropy by} & \textbf{Standard} & \textbf{Entropy by} \\
& \textbf{brand} & \textbf{deviation of} & \textbf{brands' price} \\
& & \textbf{brands' SES} & \textbf{level} \\
\midrule
Income & 0.726*** & 0.085*** & 0.127*** \\
& (0.014) & (0.014) & (0.015) \\
Proportion of & 0.144*** & 0.498*** & 0.461*** \\
bachelor's degree & (0.011) & (0.011) & (0.012) \\
or higher & & & \\
Income-education & -0.218*** & 0.064*** & -0.014 \\
interaction & (0.007) & (0.007) & (0.007) \\
Income variability & -0.173*** & 0.005 & -0.003 \\
& (0.009) & (0.009) & (0.010) \\
Mobility & -0.186*** & -0.221*** & -0.063*** \\
& (0.008) & (0.009) & (0.009) \\
Local availability & 0.140*** & 0.110*** & 0.110*** \\
& (0.008) & (0.008) & (0.008) \\
In Greater Houston & 0.274*** & 0.436*** & 0.307*** \\
& (0.019) & (0.019) & (0.021) \\
Median age & -0.103*** & -0.012 & 0.024** \\
& (0.008) & (0.008) & (0.009) \\
Proportion of male & -0.029*** & 0.017* & 0.015 \\
& (0.007) & (0.007) & (0.008) \\
Proportion of white & -0.050*** & -0.093*** & 0.014 \\
& (0.008) & (0.008) & (0.009) \\
Intercept & 0.099*** & -0.131*** & -0.051*** \\
& (0.010) & (0.010) & (0.011) \\
\midrule
Observations & 9571 & 7202 & 9007 \\
R² & 0.479 & 0.611 & 0.375 \\
Adjusted R² & 0.479 & 0.610 & 0.374 \\
Residual Std. Error & 0.722 & 0.628 & 0.784 \\
F Statistic & 879.770*** & 1128.589*** & 540.136*** \\
\bottomrule
\end{tabular}
\begin{flushleft}
\footnotesize
Note: * $p < 0.05$, ** $p < 0.01$, *** $p < 0.001$ (two-tailed tests).
\end{flushleft}
\end{table}

\begin{table}[ht]
\centering
\caption{Results from regression analyses that predict CBGs' diversity (entropy by brand) by industry (Texas).}
\label{tab:tab_f6}
\tiny
\setlength{\tabcolsep}{3pt}
\begin{tabular}{lrrrrrrrrr}
\toprule
& \textbf{Amusement,} & \textbf{Clothing and} & \textbf{Food Services} & \textbf{Food and} & \textbf{Gasoline} & \textbf{General} & \textbf{Health and} & \textbf{Miscellaneous} & \textbf{Sporting} \\
& \textbf{Gambling,} & \textbf{Clothing} & \textbf{and Drinking} & \textbf{Beverage} & \textbf{Stations} & \textbf{Merchandise} & \textbf{Personal Care} & \textbf{Store} & \textbf{Goods,} \\
& \textbf{and Recreation} & \textbf{Accessories} & \textbf{Places} & \textbf{Stores} & & \textbf{Stores} & \textbf{Stores} & \textbf{Retailers} & \textbf{Hobby} \\
& \textbf{Industries} & \textbf{Stores} & & & & & & & \\
\midrule
Income & 0.669*** & 0.524*** & 0.693*** & 0.333*** & 0.569*** & 0.354*** & 0.434*** & 0.547*** & 0.496*** \\
& (0.032) & (0.054) & (0.017) & (0.025) & (0.020) & (0.028) & (0.033) & (0.040) & (0.051) \\
Proportion of & 0.156*** & 0.052 & 0.139*** & 0.001 & -0.142*** & 0.152*** & 0.023 & 0.049 & 0.114** \\
bachelor's degree & (0.023) & (0.039) & (0.013) & (0.019) & (0.017) & (0.024) & (0.025) & (0.029) & (0.038) \\
or higher & & & & & & & & & \\
Income-education & -0.192*** & -0.126*** & -0.215*** & -0.084*** & -0.209*** & -0.107*** & -0.132*** & -0.144*** & -0.138*** \\
interaction & (0.015) & (0.027) & (0.008) & (0.012) & (0.011) & (0.016) & (0.016) & (0.018) & (0.025) \\
Income variability & -0.175*** & -0.173*** & -0.177*** & -0.056*** & -0.115*** & -0.115*** & -0.132*** & -0.162*** & -0.169*** \\
& (0.018) & (0.033) & (0.011) & (0.016) & (0.013) & (0.018) & (0.021) & (0.022) & (0.031) \\
Mobility & 0.020 & -0.027 & -0.176*** & 0.022 & 0.103*** & -0.190*** & 0.038 & 0.236*** & -0.098** \\
& (0.033) & (0.032) & (0.011) & (0.016) & (0.012) & (0.016) & (0.031) & (0.054) & (0.034) \\
Local availability & 0.143*** & 0.280*** & 0.152*** & 0.048*** & 0.148*** & 0.110*** & 0.147*** & 0.131*** & 0.135*** \\
& (0.018) & (0.030) & (0.010) & (0.013) & (0.011) & (0.015) & (0.019) & (0.022) & (0.028) \\
In Greater Houston & 0.135** & 0.410*** & 0.056* & 0.684*** & 0.400*** & 0.265*** & 0.279*** & 0.184*** & 0.094 \\
& (0.042) & (0.072) & (0.024) & (0.035) & (0.028) & (0.039) & (0.045) & (0.050) & (0.068) \\
Median age & -0.135*** & -0.195*** & -0.101*** & -0.097*** & 0.002 & -0.114*** & -0.091*** & -0.133*** & -0.116*** \\
& (0.021) & (0.033) & (0.011) & (0.015) & (0.012) & (0.015) & (0.022) & (0.026) & (0.032) \\
Proportion of male & -0.039* & -0.064* & -0.031*** & -0.007 & -0.042*** & -0.007 & -0.084*** & -0.070** & -0.067* \\
& (0.019) & (0.029) & (0.009) & (0.014) & (0.011) & (0.015) & (0.020) & (0.023) & (0.030) \\
Proportion of white & -0.073*** & -0.020 & -0.052*** & -0.220*** & -0.145*** & 0.008 & -0.029 & -0.061* & -0.045 \\
& (0.022) & (0.033) & (0.010) & (0.015) & (0.012) & (0.015) & (0.021) & (0.025) & (0.032) \\
Intercept & 0.369*** & 0.196*** & 0.196*** & -0.042* & 0.127*** & 0.062** & 0.171*** & 0.406*** & 0.286*** \\
& (0.024) & (0.039) & (0.012) & (0.018) & (0.014) & (0.019) & (0.024) & (0.033) & (0.037) \\
\midrule
Observations & 1595 & 818 & 6010 & 3671 & 6214 & 3509 & 2053 & 1160 & 750 \\
R² & 0.434 & 0.322 & 0.429 & 0.256 & 0.210 & 0.212 & 0.177 & 0.286 & 0.269 \\
Adjusted R² & 0.430 & 0.314 & 0.428 & 0.254 & 0.209 & 0.210 & 0.173 & 0.280 & 0.259 \\
Residual Std. Error & 0.691 & 0.830 & 0.738 & 0.809 & 0.864 & 0.844 & 0.838 & 0.725 & 0.752 \\
F Statistic & 121.389*** & 38.366*** & 451.107*** & 126.212*** & 164.680*** & 94.320*** & 43.836*** & 46.000*** & 27.230*** \\
\bottomrule
\end{tabular}
\begin{flushleft}
\footnotesize
Note: *** $p < 0.001$, ** $p < 0.01$, * $p < 0.05$ (two-tailed tests). Excluded the industry \emph{Motion Picture and Video Industries} due to limited number of observations with price and SES data. Excluded \emph{Personal and Laundry Services} due to limited number of observations.
\end{flushleft}
\end{table}

\begin{table}[ht]
\centering
\caption{Results from regression analyses that predict CBGs' diversity (standard deviation of brands' SES) by industry (Texas).}
\label{tab:tab_f7}
\tiny
\setlength{\tabcolsep}{3pt}
\begin{tabular}{lrrrrrrrrr}
\toprule
& \textbf{Amusement,} & \textbf{Clothing and} & \textbf{Food Services} & \textbf{Food and} & \textbf{Gasoline} & \textbf{General} & \textbf{Health and} & \textbf{Miscellaneous} & \textbf{Sporting} \\
& \textbf{Gambling,} & \textbf{Clothing} & \textbf{and Drinking} & \textbf{Beverage} & \textbf{Stations} & \textbf{Merchandise} & \textbf{Personal Care} & \textbf{Store} & \textbf{Goods,} \\
& \textbf{and Recreation} & \textbf{Accessories} & \textbf{Places} & \textbf{Stores} & & \textbf{Stores} & \textbf{Stores} & \textbf{Retailers} & \textbf{Hobby} \\
& \textbf{Industries} & \textbf{Stores} & & & & & & & \\
\midrule
Income & 0.129* & 0.118 & 0.078*** & 0.019 & 0.231*** & 0.233*** & 0.224*** & 0.174* & -0.194* \\
& (0.064) & (0.078) & (0.020) & (0.045) & (0.027) & (0.042) & (0.067) & (0.080) & (0.094) \\
Proportion of & 0.118** & 0.106 & 0.484*** & 0.172*** & 0.224*** & 0.351*** & 0.056 & 0.177** & 0.128* \\
bachelor's degree & (0.042) & (0.056) & (0.015) & (0.034) & (0.023) & (0.036) & (0.053) & (0.062) & (0.064) \\
or higher & & & & & & & & & \\
Income-education & -0.032 & -0.015 & 0.057*** & 0.018 & -0.038* & 0.011 & -0.060 & -0.021 & 0.220*** \\
interaction & (0.031) & (0.035) & (0.010) & (0.023) & (0.015) & (0.023) & (0.034) & (0.039) & (0.042) \\
Income variability & -0.051 & -0.014 & 0.024 & 0.036 & -0.032 & -0.049 & -0.096* & 0.023 & -0.011 \\
& (0.035) & (0.040) & (0.013) & (0.029) & (0.017) & (0.027) & (0.041) & (0.045) & (0.047) \\
Mobility & 0.101 & -0.125* & -0.229*** & -0.067 & -0.075*** & -0.177*** & 0.009 & 0.074 & -0.060 \\
& (0.116) & (0.054) & (0.015) & (0.038) & (0.017) & (0.027) & (0.071) & (0.127) & (0.068) \\
Local availability & 0.141*** & 0.012 & 0.134*** & 0.215*** & 0.231*** & 0.117*** & 0.244*** & 0.203*** & 0.357*** \\
& (0.032) & (0.043) & (0.012) & (0.025) & (0.014) & (0.022) & (0.038) & (0.043) & (0.046) \\
In Greater Houston & -0.107 & 0.080 & 0.290*** & 0.386*** & 0.759*** & 0.172** & 0.711*** & -0.105 & -0.030 \\
& (0.073) & (0.097) & (0.027) & (0.060) & (0.034) & (0.055) & (0.087) & (0.093) & (0.109) \\
Median age & 0.004 & -0.071 & 0.009 & -0.072** & 0.030 & 0.027 & 0.042 & 0.029 & 0.021 \\
& (0.045) & (0.047) & (0.013) & (0.027) & (0.016) & (0.024) & (0.049) & (0.055) & (0.060) \\
Proportion of male & -0.029 & -0.075 & 0.014 & -0.009 & -0.039** & -0.033 & -0.068 & -0.136** & -0.066 \\
& (0.038) & (0.043) & (0.011) & (0.025) & (0.014) & (0.023) & (0.042) & (0.048) & (0.056) \\
Proportion of white & 0.059 & -0.058 & -0.094*** & -0.173*** & -0.053** & -0.103*** & -0.012 & -0.133** & -0.102 \\
& (0.041) & (0.046) & (0.012) & (0.028) & (0.016) & (0.023) & (0.045) & (0.047) & (0.053) \\
Intercept & 0.407*** & 0.119* & -0.072*** & -0.109*** & -0.142*** & -0.100*** & 0.178*** & 0.254*** & 0.096 \\
& (0.061) & (0.056) & (0.015) & (0.032) & (0.019) & (0.028) & (0.051) & (0.071) & (0.064) \\
\midrule
Observations & 373 & 364 & 3822 & 1007 & 2510 & 957 & 540 & 356 & 241 \\
R² & 0.155 & 0.097 & 0.559 & 0.272 & 0.427 & 0.434 & 0.258 & 0.258 & 0.400 \\
Adjusted R² & 0.132 & 0.071 & 0.557 & 0.264 & 0.425 & 0.428 & 0.244 & 0.237 & 0.374 \\
Residual Std. Error & 0.611 & 0.775 & 0.686 & 0.753 & 0.713 & 0.653 & 0.860 & 0.766 & 0.688 \\
F Statistic & 6.664*** & 3.771*** & 482.304*** & 37.156*** & 186.485*** & 72.492*** & 18.400*** & 12.015*** & 15.354*** \\
\bottomrule
\end{tabular}
\begin{flushleft}
\footnotesize
Note: *** $p < 0.001$, ** $p < 0.01$, * $p < 0.05$ (two-tailed tests). Excluded the industry \emph{Motion Picture and Video Industries} due to limited number of observations with price and SES data. Excluded \emph{Personal and Laundry Services} due to limited number of observations.
\end{flushleft}
\end{table}

\begin{table}[ht]
\centering
\caption{Results from regression analyses that predict CBGs' diversity (entropy by brands' price level) by industry (Texas).}
\label{tab:tab_f8}
\tiny
\setlength{\tabcolsep}{3pt}
\begin{tabular}{lrrrrrrrrr}
\toprule
& \textbf{Amusement,} & \textbf{Clothing and} & \textbf{Food Services} & \textbf{Food and} & \textbf{Gasoline} & \textbf{General} & \textbf{Health and} & \textbf{Miscellaneous} & \textbf{Sporting} \\
& \textbf{Gambling,} & \textbf{Clothing} & \textbf{and Drinking} & \textbf{Beverage} & \textbf{Stations} & \textbf{Merchandise} & \textbf{Personal Care} & \textbf{Store} & \textbf{Goods,} \\
& \textbf{and Recreation} & \textbf{Accessories} & \textbf{Places} & \textbf{Stores} & & \textbf{Stores} & \textbf{Stores} & \textbf{Retailers} & \textbf{Hobby} \\
& \textbf{Industries} & \textbf{Stores} & & & & & & & \\
\midrule
Income & 0.566*** & 0.237*** & 0.198*** & 0.119*** & 0.182*** & 0.296*** & 0.043 & 0.268*** & -0.003 \\
& (0.072) & (0.065) & (0.016) & (0.029) & (0.024) & (0.026) & (0.041) & (0.050) & (0.069) \\
Proportion of & 0.076 & 0.127** & 0.515*** & 0.144*** & -0.051* & 0.448*** & 0.041 & -0.061 & 0.209*** \\
bachelor's degree & (0.058) & (0.044) & (0.013) & (0.022) & (0.020) & (0.023) & (0.031) & (0.036) & (0.052) \\
or higher & & & & & & & & & \\
Income-education & -0.221*** & -0.035 & -0.051*** & -0.076*** & -0.082*** & -0.122*** & -0.025 & -0.072** & -0.008 \\
interaction & (0.039) & (0.031) & (0.008) & (0.014) & (0.013) & (0.015) & (0.019) & (0.023) & (0.035) \\
Income variability & -0.207*** & -0.048 & -0.027** & 0.003 & -0.007 & -0.067*** & -0.050* & -0.107*** & -0.036 \\
& (0.047) & (0.037) & (0.010) & (0.018) & (0.016) & (0.017) & (0.025) & (0.028) & (0.043) \\
Mobility & -0.048 & 0.027 & -0.131*** & -0.058** & 0.057*** & -0.164*** & 0.019 & 0.243*** & -0.028 \\
& (0.070) & (0.038) & (0.011) & (0.018) & (0.014) & (0.015) & (0.042) & (0.066) & (0.045) \\
Local availability & 0.086* & 0.201*** & 0.126*** & 0.085*** & 0.118*** & 0.107*** & 0.163*** & 0.092*** & 0.181*** \\
& (0.038) & (0.035) & (0.009) & (0.016) & (0.013) & (0.014) & (0.023) & (0.026) & (0.038) \\
In Greater Houston & -0.076 & 0.399*** & 0.250*** & -0.028 & 0.205*** & 0.130*** & 0.164** & 0.177** & -0.253** \\
& (0.109) & (0.082) & (0.022) & (0.040) & (0.032) & (0.037) & (0.053) & (0.061) & (0.092) \\
Median age & -0.116** & -0.066 & 0.010 & -0.071*** & -0.007 & -0.094*** & -0.013 & -0.092** & -0.055 \\
& (0.044) & (0.040) & (0.010) & (0.018) & (0.014) & (0.015) & (0.027) & (0.032) & (0.045) \\
Proportion of male & -0.050 & -0.012 & 0.013 & 0.019 & -0.023 & 0.009 & -0.050* & -0.014 & 0.078 \\
& (0.042) & (0.033) & (0.009) & (0.016) & (0.013) & (0.014) & (0.024) & (0.027) & (0.041) \\
Proportion of white & 0.017 & -0.045 & -0.038*** & -0.116*** & -0.076*** & 0.003 & -0.020 & 0.004 & -0.077 \\
& (0.048) & (0.039) & (0.010) & (0.017) & (0.014) & (0.015) & (0.025) & (0.030) & (0.043) \\
Intercept & 0.313*** & 0.049 & -0.003 & 0.140*** & 0.066*** & -0.005 & 0.049 & 0.298*** & 0.195*** \\
& (0.050) & (0.046) & (0.011) & (0.021) & (0.017) & (0.018) & (0.030) & (0.040) & (0.049) \\
\midrule
Observations & 528 & 650 & 5949 & 3309 & 4863 & 3473 & 1788 & 982 & 703 \\
R² & 0.193 & 0.198 & 0.526 & 0.081 & 0.052 & 0.338 & 0.040 & 0.077 & 0.093 \\
Adjusted R² & 0.177 & 0.185 & 0.525 & 0.078 & 0.050 & 0.336 & 0.035 & 0.068 & 0.080 \\
Residual Std. Error & 0.882 & 0.859 & 0.688 & 0.894 & 0.897 & 0.802 & 0.954 & 0.812 & 0.984 \\
F Statistic & 12.366*** & 15.758*** & 657.878*** & 29.159*** & 26.791*** & 176.561*** & 7.462*** & 8.156*** & 7.132*** \\
\bottomrule
\end{tabular}
\begin{flushleft}
\footnotesize
Note: *** $p < 0.001$, ** $p < 0.01$, * $p < 0.05$ (two-tailed tests). Excluded the industry \emph{Motion Picture and Video Industries} due to limited number of observations with price and SES data. Excluded \emph{Personal and Laundry Services} due to limited number of observations.
\end{flushleft}
\end{table}

\clearpage
\subsection{Niche consumption}

\begin{figure}[ht]
\centering
\includegraphics[width=\textwidth]{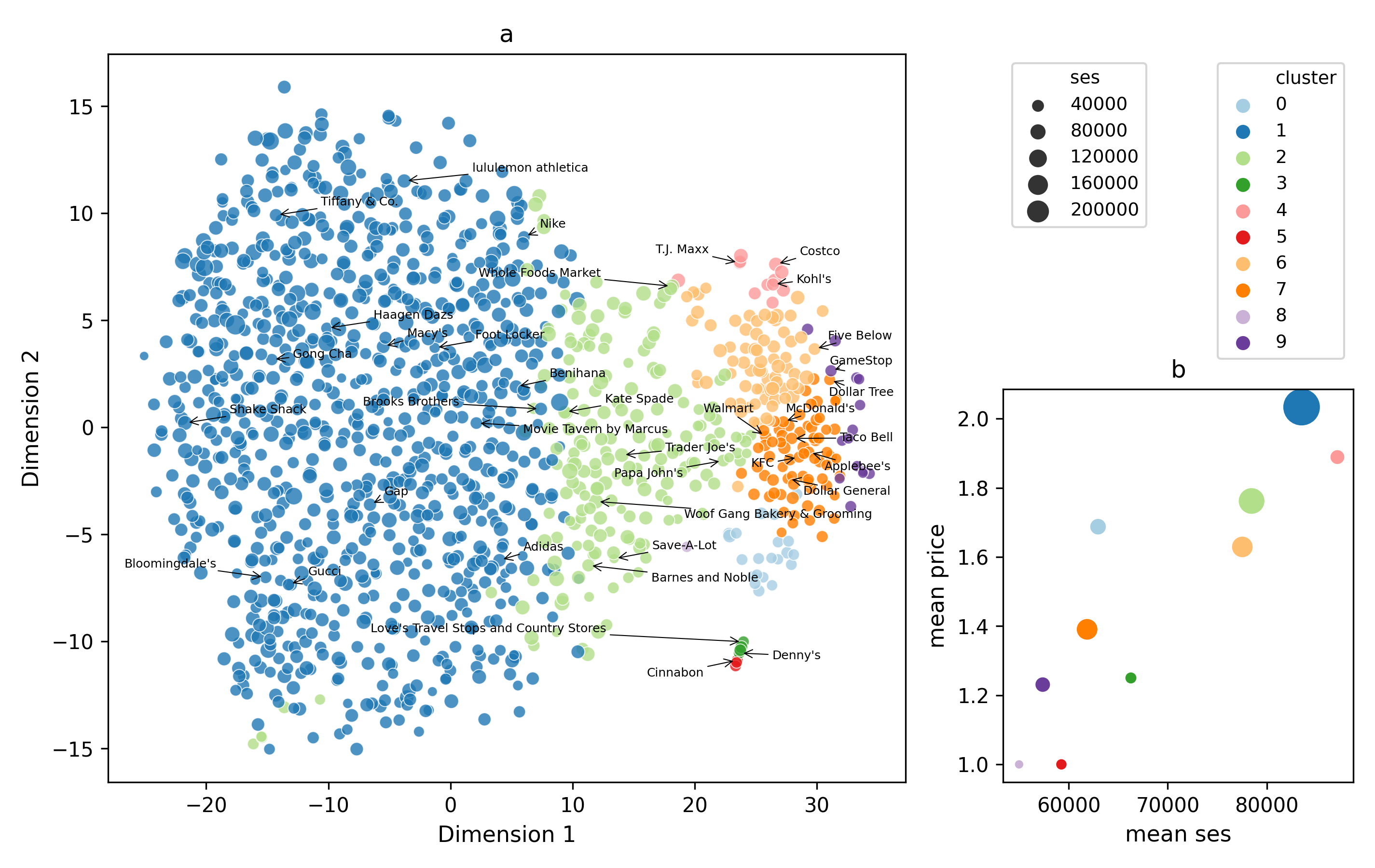}
\caption{Niche consumption analysis (Texas): a) t-SNE visualization and $k$-means clustering of the brand co-visit network after node2vec embedding to 128 dimensions; b) mean SES and price for the brands in each cluster. In panel a, the marker size represents the brand's SES, as the legend indicates. In panel b, the marker size is proportional to the square root of the number of brands in each cluster.}
\label{fig:fig_f10}
\end{figure}

We use the same process as for New York State to test niche consumption patterns with the data in Texas on brand co-visits for the 1,273 brands. The results are presented in Figure \ref{fig:fig_f10}. As in New York State, we find no evidence for the niche consumption hypothesis. Apart from few exceptions (e.g., the Cinnabon cluster), the t-SNE plot does not identify distinct consumption niches. Mapping the 10 clusters identified by the k-means algorithm by mean SES and price, we find large consumer clusters for middle and upper-middle SES and more numerous and smaller clusters for low SES. These patterns are the opposite to the expected patterns if niche consumption drives the diverse consumption in high-SES groups.

\end{document}